%% file: main.tex
\documentclass[10pt,conference]{IEEEtran} 
\IEEEoverridecommandlockouts

\usepackage{cite}
\usepackage{amsmath,amssymb,amsfonts}
\usepackage{algorithmic}
\usepackage{graphicx}
\usepackage{xcolor}
\usepackage{adjustbox}
\usepackage{amsmath}
\usepackage{tikz}
\usetikzlibrary{tikzmark}
\usepackage{quantikz}
\usepackage{caption}
\usepackage{subcaption}
\usepackage{multicol}
\usepackage{multirow}
\usepackage{nicefrac}
\usepackage{braket}
\usepackage[linesnumbered,ruled,vlined]{algorithm2e}
\usepackage{diagbox}
\usepackage{tabularx}
\usepackage{placeins}
\usepackage{nicematrix}
\usepackage{booktabs}
\usepackage{amsmath}
\usepackage{colortbl}
\usepackage{arydshln}
\usepackage[utf8]{inputenc} 
\usepackage[T1]{fontenc}    
\usepackage{hyperref}       
\usepackage{url}            
\usepackage{booktabs}       
\usepackage{amsfonts}       
\usepackage{braket}
\usepackage{tikz}
\usetikzlibrary{matrix,decorations.pathreplacing}
\usetikzlibrary{matrix,calc}
\usepackage{todonotes}
\usepackage{nicefrac}       
\usepackage{microtype}      
\usepackage{lipsum}
\usepackage{fancyhdr}       
\usepackage{graphicx}       
\usepackage{adjustbox}
\usepackage{comment}
\usepackage[numbers]{natbib}
\setcitestyle{numbers,square,comma,aysep={},yysep={,}}
\setcitestyle{comma,numbers,square,separated}
\usepackage{csvsimple}
\usepackage{arydshln}

\input{tikzpreamble}

\title{Redefining Lexicographical Ordering: Optimizing Pauli String Decompositions for Quantum Compiling}

\author{\IEEEauthorblockN{Qunsheng Huang, David Winderl}
\IEEEauthorblockA{\textit{CIT Department of Computer Science} \\
\textit{Technical University of Munich}\\
Garching, Germany \\
keefe.huang@tum.de, david.winderl@tum.de}
\and
\IEEEauthorblockN{Arianne Meijer--van de Griend}
\IEEEauthorblockA{\textit{Department of Computer Science} \\
\textit{University of Helsinki}\\
Helsinki, Finland \\
ariannemeijer@gmail.com}
\and
\IEEEauthorblockN{Richie Yeung}
\IEEEauthorblockA{\textit{Department of Computer Science} \\
\textit{University of Oxford}\\
Oxford, United Kingdom \\
richie.yeung@cs.ox.ac.uk}
\and
\IEEEauthorblockA{All authors contributed equally.}
}

\tikzset{OPLUS/.style={draw,circle,append after command={
        [shorten >=\pgflinewidth, shorten <=\pgflinewidth,]
        (\tikzlastnode.north) edge (\tikzlastnode.south)
        (\tikzlastnode.east) edge (\tikzlastnode.west)
        }
    }
}
\tikzset{OMINUS/.style={draw,circle,append after command={
        [shorten >=\pgflinewidth, shorten <=\pgflinewidth,]
        (\tikzlastnode.east) edge (\tikzlastnode.west)
        }
    }
}

\newcommand*{\leftarrowpm}[1]{%
  \xleftarrow{\scriptscriptstyle\!#1\!}%
}

\newcommand{\tn}[1]{
    \text{#1}%
}

\renewcommand{\bf}[1]{\mathbf{#1}}

\begin{document}

\maketitle







\begin{abstract}
In quantum computing, the efficient optimization of Pauli string decompositions is a crucial aspect for the compilation of quantum circuits for many applications, such as chemistry simulations and quantum machine learning. In this paper, we propose a novel algorithm for the synthesis of trotterized time-evolution operators that results in circuits with significantly fewer gates than previous solutions. Our synthesis procedure takes the qubit connectivity of a target quantum computer into account. As a result, the generated quantum circuit does not require routing, and no additional CNOT gates are needed to run the resulting circuit on a target device. We compare our algorithm against Paulihedral and TKET, and show a significant improvement for randomized circuits and different molecular ansatzes. We also investigate the Trotter error introduced by our ordering of the terms in the Hamiltonian versus default ordering and the ordering from the baseline methods and conclude that our method on average does not increase the Trotter error.
\end{abstract}


\section{Introduction}
\input{sections/introduction}

\section{Preliminaries and Related Work}
\input{sections/preliminaries}

\input{sections/related_work}

\section{Methods}
\input{sections/methods}

\section{Experiments}
\input{sections/experiments}

\section{Conclusion}
\input{sections/conclusion}

\FloatBarrier
\clearpage

\newpage

\bibliographystyle{IEEEtran}

\bibliography{references}

\end{document}

%% file: tikzpreamble.tex
\usepackage{tikzit}  

\newcommand{\inlinetikzfig}[2][1.0]{
  \dimendef\grafheight=255\dimendef\grafvshift=254
  \setbox0 = \hbox{\scalebox{#1}{\input{\tikzitpath#2.tikz}}}
  \grafheight=\the\ht0
  \grafvshift=-0.5\grafheight
  \advance\grafvshift by 0.5ex
  \raisebox{\grafvshift}{\input{\tikzitpath#2.tikz}}
}

\pgfdeclarelayer{background} 
\pgfdeclarelayer{foreground}
\pgfdeclarelayer{edgelayer}
\pgfdeclarelayer{nodelayer}
\pgfsetlayers{background,edgelayer,nodelayer,main,foreground} 

\tikzstyle{halfsize}=[x=0.5cm, y=0.5cm]
\tikzstyle{normalsize}=[]
\tikzstyle{doublesize}=[]

\tikzstyle{(null)}=[]
\tikzstyle{plain}=[]

\usepackage{tikzzx}
\usepackage{tikzcircuits}
\usepackage[undirected]{tikzquanto} 
\usetikzlibrary{calc}

\tikzset{every picture/.style={line width=0.75pt}} 
\tikzset{directed edge/.style={postaction={decorate,decoration={markings,mark=at position 0.5 with {\arrow{>}}}}}}

%% file: sections/introduction.tex
With the increasing size and quality of current quantum computers comes a need for efficient compilation algorithms that can process more complicated circuits in polynomial time without generating more operations than necessary. In this paper, we focus on the synthesis of Hamiltonian time evolution operators to quantum circuits constrained under the qubit connectivity of a target quantum computer. These operators can be found in the ubiquitous Ising model as well as algorithms such as the unitary-coupled cluster ansatz (UCC) \cite{Anand2022} or the Harrow–Hassidim–Lloyd (HHL) algorithm \cite{Harrow2009}, among others. Similarly, sequences of multiple controlled rotations can also be represented as a sequence of Pauli string operations which are common in e.g. Amplitude Encoding~\cite{Nielsen2012}, and Linear Combinations of Unitaries~\cite{childs2012hamiltonian}, among others.
Moreover, they form a universal representation, so any quantum algorithm can be rewritten into a product formula that contains only time evolution operators. 
While time evolution operators in algorithms can be easily determined for classical simulation, transpiling the same operations to gates on a quantum device is not obvious. 

In this context, our proposed method synthesizes a quantum circuit from the first-order Suzuki-Trotter decomposition of the target Hamiltonian's time evolution operator (i.e. a product of exponentiated Pauli strings). Additionally, the algorithm takes qubit connectivity into account during synthesis. As a result, the synthesized circuit is already routed and no additional SWAP gates are required to run the circuit on the target device.  
Along the way, we extend a previous method of ours~\cite{winderl2023architectureaware} to more effectively synthesize Clifford tableaus by allowing a final permutation either implemented on the quantum circuit or by a permutation of swaps.

Our experiments show that the introduced algorithm significantly improves the number of CNOTs, the circuit depth, and the CNOT depth with respect to Paulihedral~\cite{li2022paulihedral} and TKET~\cite{cowtan2020generic}, both in the constrained and unconstrained case. In the case of targeting a restricted architecture, our proposed method improves these metrics up to an order of magnitude for UCCSD ansatzes.

%% file: sections/preliminaries.tex
\subsection{Pauli Gadgets and Pauli Polynomials}
\label{subsec:phase_gadgets}
In this work, we consider Trotterized time evolution operators written in the form \begin{equation}\label{eq:trotter}
   U = e^{-i \sum_n \left(\frac{\theta_n}{2} \bigotimes_m P_m \right)} \approx \left(\prod_n e^{-i \frac{\theta_n}{2r} \bigotimes_m P_m}\right)^r,
\end{equation}
where $P_m$ is a single-qubit Pauli operator, and $r$ the number of repetitions of the circuit.  The tensor of Pauli gates can also be written as a sequence of the letters representing each Pauli gate i.e. representing $X \otimes I \otimes X \otimes Y \otimes Z$ as $XIXYZ$. This sequence is also called a Pauli string because it is a word $s$ of arbitrary length over the alphabet $\{I, X, Y, Z\}$, or the four Pauli \emph{letters}. 

Following the conventions in \cite{cowtan2019phase} and \cite{yeung2020diagrammatic}, we define a Pauli gadget $\phi$ with a Pauli string $s$ and angle $\theta$:
\begin{equation}
    \phi(\theta, s) = \exp \left(-i\frac{\theta}{2} s \right)
\end{equation}
Additionally, we also utilize their convenient graphical representation of the Pauli gadget in the ZX-calculus:
\begin{equation}\label{eq:phase_gadget}
    \exp\left(-i\frac{\theta}{2}IXYZ\right) = \input{\tikzitpathfigures/PauliExpDef.tikz}
\end{equation}

 In this representation, each non-identity Pauli letter in $s$ maps to a \emph{leg} on the corresponding qubit or wire, each attached to a central "spider" containing the angle information, as shown in \autoref{eq:phase_gadget}.
 Then, given a Hamiltonian described by a sum of weighted Pauli strings, we can approximate the corresponding time evolution operator in terms of Pauli gadgets. We refer to such a representation as a \emph{Pauli polynomial}. For example, given one such arbitrary Hamiltonian $H = \alpha IXYX + \beta IZZI + \alpha IXYY + \beta IYYI$, where $\alpha, \beta \in \mathbb{R}$, we could approximate the time evolution operator over a time step $t$ as follows:
\begin{equation}\label{eq:pauli_polynomial}
    \input{\tikzitpathfigures/PauliExpTrotter.tikz}
\end{equation}
This notation is equivalent to that used in \cite{li2022paulihedral} with the difference being that our notation can be considered a dialect of the ZX-calculus~\cite{wetering2020zx} and, therefore, all equalities presented in this work can be considered mathematical equations. 
Additionally, this representation includes well-defined rules for propagating Clifford gates through Pauli gadgets \cite{cowtan2020generic,yeung2020diagrammatic} that are key for the proposed methods.
The graphical representations of the Clifford operators relevant to this work are then as follows:
\begin{align}
V = \input{\tikzitpathfigures/VDef.tikz}  \quad
V^\dagger &= \input{\tikzitpathfigures/VDDef.tikz} \quad
H = \input{\tikzitpathfigures/HDef.tikz} \quad
CNOT = \input{\tikzitpathfigures/CNOTDef.tikz} \\
S &= \input{\tikzitpathfigures/SDef.tikz} \quad S^\dagger = \input{\tikzitpathfigures/SDagDef.tikz} ,\nonumber
\end{align}
where $V := \sqrt{X}$.
Then, propagating a CNOT through gadgets with $ZZ$-legs or $ZY$-legs is represented as follows:
\begin{equation}\label{eq:CXProp}
{\input{\tikzitpathfigures/PauliExpCXZZra.tikz}} = {\input{\tikzitpathfigures/PauliExpCXZZrb.tikz}}
\end{equation}

This equality is used to remove legs from Pauli gadgets with a set formulation in the proposed method. The remaining propagation rules used in this work to manipulate legs are:

\begin{equation}\label{eq:VProp}
{\input{\tikzitpathfigures/PauliPropva.tikz}} = {\input{\tikzitpathfigures/PauliPropvb.tikz}}
\end{equation}
\begin{equation}\label{eq:HProp}
{\input{\tikzitpathfigures/PauliPropha.tikz}} = {\input{\tikzitpathfigures/PauliProphb.tikz}}
\end{equation}
\begin{equation}\label{eq:SProp}
{\input{\tikzitpathfigures/PauliPropsa.tikz}} = {\input{\tikzitpathfigures/PauliPropsb.tikz}}
\end{equation}

%% file: figures/PauliExpDef.tikz
\begin{tikzpicture}[baseline={(13)}]
	\begin{pgfonlayer}{nodelayer}
		\node [style=none] (0) at (0, 0) {};
		\node [style=none] (1) at (0, 0.75) {};
		\node [style=none] (2) at (0, 1.5) {};
		\node [style=none] (3) at (0, 2.25) {};
		\node [style=Zexp] (4) at (0.75, 0) {$Z$};
		\node [style=Xexp] (5) at (0.75, 1.5) {$X$};
		\node [style=Yexp] (6) at (0.75, 0.75) {$Y$};
		\node [style=none] (9) at (2.5, 2.25) {};
		\node [style=none] (10) at (2.5, 1.5) {};
		\node [style=none] (11) at (2.5, 0.75) {};
		\node [style=none] (12) at (2.5, 0) {};
		\node [style=Aexp] (13) at (1.75, 1.125) {$\theta$};
	\end{pgfonlayer}
	\begin{pgfonlayer}{edgelayer}
		\draw (3.center) to (9.center);
		\draw (2.center) to (5);
		\draw (5) to (10.center);
		\draw (6) to (1.center);
		\draw (6) to (11.center);
		\draw (12.center) to (4);
		\draw (4) to (0.center);
		\draw (5) to (13.center);
		\draw (13) to (6);
		\draw (4) to (13.center);
	\end{pgfonlayer}
\end{tikzpicture}

%% file: figures/PauliExpTrotter.tikz
\begin{tikzpicture}[baseline={(13)}]
	\begin{pgfonlayer}{nodelayer}
		\node [style=none] (0) at (0, 0) {};
		\node [style=none] (1) at (0, 0.75) {};
		\node [style=none] (2) at (0, 1.5) {};
		\node [style=none] (3) at (0, 2.25) {};
		\node [style=Xexp] (4) at (0.75, 0) {};
		\node [style=Xexp] (5) at (0.75, 1.5) {};
		\node [style=Yexp] (6) at (0.75, 0.75) {};
		\node [style=none] (9) at (2.5, 2.25) {};
		\node [style=none] (10) at (2.5, 1.5) {};
		\node [style=none] (11) at (2.5, 0.75) {};
		\node [style=none] (12) at (2.5, 0) {};
		\node [style=Aexp] (13) at (1.75, 1.125) {$\alpha t$};
		\node [style=none] (14) at (2.5, 0) {};
		\node [style=none] (15) at (2.5, 0.75) {};
		\node [style=none] (16) at (2.5, 1.5) {};
		\node [style=none] (17) at (2.5, 2.25) {};
		\node [style=Zexp] (19) at (3.25, 1.5) {};
		\node [style=Zexp] (20) at (3.25, 0.75) {};
		\node [style=none] (21) at (5, 2.25) {};
		\node [style=none] (22) at (5, 1.5) {};
		\node [style=none] (23) at (5, 0.75) {};
		\node [style=none] (24) at (5, 0) {};
		\node [style=Aexp] (25) at (4.25, 1.125) {$\beta t$};
		\node [style=none] (26) at (5, 0) {};
		\node [style=none] (27) at (5, 0.75) {};
		\node [style=none] (28) at (5, 1.5) {};
		\node [style=none] (29) at (5, 2.25) {};
		\node [style=Yexp] (30) at (5.75, 0) {};
		\node [style=Xexp] (31) at (5.75, 1.5) {};
		\node [style=Yexp] (32) at (5.75, 0.75) {};
		\node [style=none] (33) at (7.5, 2.25) {};
		\node [style=none] (34) at (7.5, 1.5) {};
		\node [style=none] (35) at (7.5, 0.75) {};
		\node [style=none] (36) at (7.5, 0) {};
		\node [style=Aexp] (37) at (6.75, 1.125) {$\alpha t$};
		\node [style=none] (38) at (7.5, 0) {};
		\node [style=none] (39) at (7.5, 0.75) {};
		\node [style=none] (40) at (7.5, 1.5) {};
		\node [style=none] (41) at (7.5, 2.25) {};
		\node [style=Yexp] (42) at (8.25, 1.5) {};
		\node [style=Yexp] (43) at (8.25, 0.75) {};
		\node [style=none] (44) at (10, 2.25) {};
		\node [style=none] (45) at (10, 1.5) {};
		\node [style=none] (46) at (10, 0.75) {};
		\node [style=none] (47) at (10, 0) {};
		\node [style=Aexp] (48) at (9.25, 1.125) {$\beta t$};
	\end{pgfonlayer}
	\begin{pgfonlayer}{edgelayer}
		\draw (3.center) to (9.center);
		\draw (2.center) to (5);
		\draw (5) to (10.center);
		\draw (6) to (1.center);
		\draw (6) to (11.center);
		\draw (12.center) to (4);
		\draw (4) to (0.center);
		\draw (5) to (13);
		\draw (13) to (6);
		\draw (4) to (13.center);
		\draw (17.center) to (21.center);
		\draw (16.center) to (19);
		\draw (19) to (22.center);
		\draw (20) to (15.center);
		\draw (20) to (23.center);
		\draw (19) to (25);
		\draw (25) to (20);
		\draw (14.center) to (24.center);
		\draw (29.center) to (33.center);
		\draw (28.center) to (31);
		\draw (31) to (34.center);
		\draw (32) to (27.center);
		\draw (32) to (35.center);
		\draw (36.center) to (30);
		\draw (30) to (26.center);
		\draw (31) to (37);
		\draw (37) to (32);
		\draw (30) to (37.center);
		\draw (41.center) to (44.center);
		\draw (40.center) to (42);
		\draw (42) to (45.center);
		\draw (43) to (39.center);
		\draw (43) to (46.center);
		\draw (42) to (48);
		\draw (48) to (43);
		\draw (38.center) to (47.center);
	\end{pgfonlayer}
\end{tikzpicture}

%% file: figures/VDef.tikz
\tikzset{OPLUS/.style={draw,circle,append after command={
        [shorten >=\pgflinewidth, shorten <=\pgflinewidth,]
        (\tikzlastnode.north) edge (\tikzlastnode.south)
        (\tikzlastnode.east) edge (\tikzlastnode.west)
        }
    }
}
\tikzset{OMINUS/.style={draw,circle,append after command={
        [shorten >=\pgflinewidth, shorten <=\pgflinewidth,]
        (\tikzlastnode.east) edge (\tikzlastnode.west)
        }
    }
}
\begin{tikzpicture}[baseline={(0, -0.05)}]
	\begin{pgfonlayer}{nodelayer}
        \node [style=none] (wire_start) at (0., 0) {};
		\node [style=none] (wire_end) at (1.0, 0) {};
        \node [style=OPLUS, fill=red] (first_X) at (0.5, 0) {};
	\end{pgfonlayer}
	\begin{pgfonlayer}{edgelayer}
        \draw (wire_start) to (wire_end);
	\end{pgfonlayer}
\end{tikzpicture}

%% file: figures/VDDef.tikz
\tikzset{OPLUS/.style={draw,circle,append after command={
        [shorten >=\pgflinewidth, shorten <=\pgflinewidth,]
        (\tikzlastnode.north) edge (\tikzlastnode.south)
        (\tikzlastnode.east) edge (\tikzlastnode.west)
        }
    }
}
\tikzset{OMINUS/.style={draw,circle,append after command={
        [shorten >=\pgflinewidth, shorten <=\pgflinewidth,]
        (\tikzlastnode.east) edge (\tikzlastnode.west)
        }
    }
}
\begin{tikzpicture}[baseline={(0, -0.05)}]
	\begin{pgfonlayer}{nodelayer}
        \node [style=none] (wire_start) at (0., 0) {};
		\node [style=none] (wire_end) at (1.0, 0) {};
        \node [style=OMINUS, fill=red] (first_X) at (0.5, 0) {};
	\end{pgfonlayer}
	\begin{pgfonlayer}{edgelayer}
        \draw (wire_start) to (wire_end);
	\end{pgfonlayer}
\end{tikzpicture}

%% file: figures/HDef.tikz
\tikzset{OPLUS/.style={draw,circle,append after command={
        [shorten >=\pgflinewidth, shorten <=\pgflinewidth,]
        (\tikzlastnode.north) edge (\tikzlastnode.south)
        (\tikzlastnode.east) edge (\tikzlastnode.west)
        }
    }
}
\tikzset{OMINUS/.style={draw,circle,append after command={
        [shorten >=\pgflinewidth, shorten <=\pgflinewidth,]
        (\tikzlastnode.east) edge (\tikzlastnode.west)
        }
    }
}
\begin{tikzpicture}[baseline={(0, -0.05)}]
	\begin{pgfonlayer}{nodelayer}
        \node [style=none] (wire_start) at (0., 0) {};
		\node [style=none] (wire_end) at (1.0, 0) {};
        \node [style=hadamard vertex] (first_X) at (0.5, 0) {};
	\end{pgfonlayer}
	\begin{pgfonlayer}{edgelayer}
        \draw (wire_start) to (wire_end);
	\end{pgfonlayer}
\end{tikzpicture}

%% file: figures/CNOTDef.tikz
\begin{tikzpicture}[baseline={(0, 0.1)}]
	\begin{pgfonlayer}{nodelayer}
        \node [style=none] (wire1_start) at (0., 0) {};
		\node [style=none] (wire1_end) at (1.0, 0) {};
        \node [style=none] (wire2_start) at (0., 0.5) {};
		\node [style=none] (wire2_end) at (1.0, 0.5) {};
  
		\node [style=red vertex] (2) at (0.5, 0) {};
		\node [style=green vertex] (3) at (0.5, 0.5) {};
	\end{pgfonlayer}
	\begin{pgfonlayer}{edgelayer}
        \draw (wire1_start) to (wire1_end);
        \draw (wire2_start) to (wire2_end);
		\draw (3) to (2);
	\end{pgfonlayer}
\end{tikzpicture}

%% file: figures/SDef.tikz
\begin{tikzpicture}[baseline={(0, -0.05)}]
	\begin{pgfonlayer}{nodelayer}
        \node [style=none] (wire_start) at (0., 0) {};
		\node [style=none] (wire_end) at (1.0, 0) {};
        \node [style=OPLUS, fill=green] (first_X) at (0.5, 0) {};
	\end{pgfonlayer}
	\begin{pgfonlayer}{edgelayer}
        \draw (wire_start) to (wire_end);
	\end{pgfonlayer}
\end{tikzpicture}

%% file: figures/SDagDef.tikz
\begin{tikzpicture}[baseline={(0, -0.05)}]
	\begin{pgfonlayer}{nodelayer}
        \node [style=none] (wire_start) at (0., 0) {};
		\node [style=none] (wire_end) at (1.0, 0) {};
        \node [style=OMINUS, fill=green] (first_X) at (0.5, 0) {};
	\end{pgfonlayer}
	\begin{pgfonlayer}{edgelayer}
        \draw (wire_start) to (wire_end);
	\end{pgfonlayer}
\end{tikzpicture}

%% file: figures/PauliExpCXZZra.tikz
\begin{tikzpicture}[baseline={(1)}]
	\begin{pgfonlayer}{nodelayer}
		\node [style=none] (0) at (-0.25, 0.5) {};
		\node [style=none] (1) at (-0.25, 0) {};
		\node [style=red vertex] (2) at (0.25, 0) {};
		\node [style=green vertex] (3) at (0.25, 0.5) {};
		\node [style=none] (7) at (1.25, -1) {};
		\node [style=none] (8) at (3.75, 0.5) {};
		\node [style=none] (9) at (3.75, 0) {};
		\node [style=none] (10) at (1.25, -0.5) {$\vdots$};
		\node [style=Zexp] (11) at (1, 0) {};
		\node [style=Aexp] (12) at (1.75, -0.5) {$\alpha$};
		\node [style=Zexp] (13) at (1, 0.5) {};
		\node [style=Zexp] (15) at (2.5, 0.5) {};
		\node [style=Yexp] (16) at (2.5, 0) {};
		\node [style=none] (17) at (2.75, -1) {};
		\node [style=none] (18) at (2.75, -0.5) {$\vdots$};
		\node [style=Aexp] (19) at (3.25, -0.5) {$\beta$};
	\end{pgfonlayer}
	\begin{pgfonlayer}{edgelayer}
		\draw (3) to (2);
		\draw (11) to (12);
		\draw (12.center) to (7.center);
		\draw (13) to (12.center);
		\draw (19.center) to (17.center);
		\draw (15) to (19.center);
		\draw (16) to (19.center);
		\draw (0.center) to (8.center);
		\draw (1.center) to (9.center);
	\end{pgfonlayer}
\end{tikzpicture}

%% file: figures/PauliExpCXZZrb.tikz
\begin{tikzpicture}[baseline={(1)}]
	\begin{pgfonlayer}{nodelayer}
		\node [style=none] (0) at (0.25, 0.5) {};
		\node [style=none] (1) at (0.25, 0) {};
		\node [style=red vertex] (2) at (3.75, 0) {};
		\node [style=green vertex] (3) at (3.75, 0.5) {};
		\node [style=none] (7) at (1.25, -1) {};
		\node [style=none] (8) at (4.25, 0.5) {};
		\node [style=none] (9) at (4.25, 0) {};
		\node [style=none] (10) at (1.25, -0.5) {$\vdots$};
		\node [style=Zexp] (11) at (1, 0) {};
		\node [style=Aexp] (12) at (1.75, -0.5) {$\alpha$};
		\node [style=Yexp] (16) at (2.5, 0) {};
		\node [style=none] (17) at (2.75, -1) {};
		\node [style=none] (18) at (2.75, -0.5) {$\vdots$};
		\node [style=Aexp] (19) at (3.25, -0.5) {$\beta$};
	\end{pgfonlayer}
	\begin{pgfonlayer}{edgelayer}
		\draw (3) to (2);
		\draw (11) to (12.center);
		\draw (12.center) to (7.center);
		\draw (19.center) to (17.center);
		\draw (16) to (19);
		\draw (1.center) to (9.center);
		\draw (0.center) to (8.center);
	\end{pgfonlayer}
\end{tikzpicture}

%% file: figures/PauliPropva.tikz
\begin{tikzpicture}[baseline={(1)}]
	\begin{pgfonlayer}{nodelayer}
		\node [style=none] (1) at (-0.25, 0) {};
		\node [style=none] (7) at (1.25, -1) {};
		\node [style=none] (10) at (1.25, -0.5) {$\vdots$};
		\node [style=Zexp] (11) at (1, 0) {};
		\node [style=Aexp] (12) at (1.75, -0.5) {$\alpha$};
		\node [style=none] (17) at (2.75, -1) {};
		\node [style=none] (18) at (2.75, -0.5) {$\vdots$};
		\node [style=Aexp] (19) at (3.25, -0.5) {$\beta$};
		\node [style=none] (20) at (5.25, 0) {};
		\node [style=none] (22) at (4.25, -1) {};
		\node [style=none] (23) at (4.25, -0.5) {$\vdots$};
		\node [style=Aexp] (24) at (4.75, -0.5) {$\delta$};
		\node [style=Xexp] (25) at (2.5, 0) {};
		\node [style=Yexp] (26) at (4, 0) {};
		\node [style=Vdag] (27) at (0.25, 0) {};
	\end{pgfonlayer}
	\begin{pgfonlayer}{edgelayer}
		\draw (11) to (12);
		\draw (12.center) to (7.center);
		\draw (19.center) to (17.center);
		\draw (1.center) to (20.center);
		\draw (24.center) to (22.center);
		\draw (25) to (19);
		\draw (26) to (24);
	\end{pgfonlayer}
\end{tikzpicture}

%% file: figures/PauliPropvb.tikz
\begin{tikzpicture}[baseline={(1)}]
	\begin{pgfonlayer}{nodelayer}
		\node [style=none] (1) at (0.5, 0) {};
		\node [style=none] (7) at (1.25, -1) {};
		\node [style=none] (10) at (1.25, -0.5) {$\vdots$};
		\node [style=Aexp] (12) at (1.75, -0.5) {$-\alpha$};
		\node [style=none] (17) at (2.75, -1) {};
		\node [style=none] (18) at (2.75, -0.5) {$\vdots$};
		\node [style=Aexp] (19) at (3.25, -0.5) {$\beta$};
		\node [style=none] (20) at (6, 0) {};
		\node [style=none] (22) at (4.25, -1) {};
		\node [style=none] (23) at (4.25, -0.5) {$\vdots$};
		\node [style=Aexp] (24) at (4.75, -0.5) {$\delta$};
		\node [style=Xexp] (25) at (2.5, 0) {};
		\node [style=Vdag] (27) at (5.5, 0) {};
		\node [style=Yexp] (28) at (1, 0) {};
		\node [style=Zexp] (29) at (4, 0) {};
	\end{pgfonlayer}
	\begin{pgfonlayer}{edgelayer}
		\draw (12.center) to (7.center);
		\draw (19.center) to (17.center);
		\draw (1.center) to (20.center);
		\draw (24.center) to (22.center);
		\draw (25) to (19);
		\draw (28) to (12);
		\draw (29) to (24);
	\end{pgfonlayer}
\end{tikzpicture}

%% file: figures/PauliPropha.tikz
\begin{tikzpicture}[baseline={(1)}]
	\begin{pgfonlayer}{nodelayer}
		\node [style=none] (1) at (-0.25, 0) {};
		\node [style=none] (7) at (1.25, -1) {};
		\node [style=none] (10) at (1.25, -0.5) {$\vdots$};
		\node [style=Zexp] (11) at (1, 0) {};
		\node [style=Aexp] (12) at (1.75, -0.5) {$\alpha$};
		\node [style=none] (17) at (2.75, -1) {};
		\node [style=none] (18) at (2.75, -0.5) {$\vdots$};
		\node [style=Aexp] (19) at (3.25, -0.5) {$\beta$};
		\node [style=none] (20) at (5.25, 0) {};
		\node [style=none] (22) at (4.25, -1) {};
		\node [style=none] (23) at (4.25, -0.5) {$\vdots$};
		\node [style=Aexp] (24) at (4.75, -0.5) {$\delta$};
		\node [style=Xexp] (25) at (2.5, 0) {};
		\node [style=Yexp] (26) at (4, 0) {};
		\node [style=hadamard vertex] (27) at (0.25, 0) {};
	\end{pgfonlayer}
	\begin{pgfonlayer}{edgelayer}
		\draw (11) to (12);
		\draw (12.center) to (7.center);
		\draw (19.center) to (17.center);
		\draw (1.center) to (20.center);
		\draw (24.center) to (22.center);
		\draw (25) to (19);
		\draw (26) to (24);
	\end{pgfonlayer}
\end{tikzpicture}

%% file: figures/PauliProphb.tikz
\begin{tikzpicture}[baseline={(1)}]
	\begin{pgfonlayer}{nodelayer}
		\node [style=none] (1) at (0.5, 0) {};
		\node [style=none] (7) at (1.25, -1) {};
		\node [style=none] (10) at (1.25, -0.5) {$\vdots$};
		\node [style=Aexp] (12) at (1.75, -0.5) {$\alpha$};
		\node [style=none] (17) at (2.75, -1) {};
		\node [style=none] (18) at (2.75, -0.5) {$\vdots$};
		\node [style=Aexp] (19) at (3.25, -0.5) {$\beta$};
		\node [style=none] (20) at (6, 0) {};
		\node [style=none] (22) at (4.25, -1) {};
		\node [style=none] (23) at (4.25, -0.5) {$\vdots$};
		\node [style=Aexp] (24) at (4.75, -0.5) {$-\delta$};
		\node [style=Yexp] (26) at (4, 0) {};
		\node [style=hadamard vertex] (27) at (5.5, 0) {};
		\node [style=Xexp] (28) at (1, 0) {};
		\node [style=Zexp] (29) at (2.5, 0) {};
	\end{pgfonlayer}
	\begin{pgfonlayer}{edgelayer}
		\draw (12.center) to (7.center);
		\draw (19.center) to (17.center);
		\draw (1.center) to (20.center);
		\draw (24.center) to (22.center);
		\draw (26) to (24);
		\draw (28) to (12);
		\draw (29) to (19);
	\end{pgfonlayer}
\end{tikzpicture}

%% file: figures/PauliPropsa.tikz
\begin{tikzpicture}[baseline={(1)}]
	\begin{pgfonlayer}{nodelayer}
		\node [style=none] (1) at (-0.25, 0) {};
		\node [style=none] (7) at (1.25, -1) {};
		\node [style=none] (10) at (1.25, -0.5) {$\vdots$};
		\node [style=Aexp] (12) at (1.75, -0.5) {$\alpha$};
		\node [style=none] (17) at (2.75, -1) {};
		\node [style=none] (18) at (2.75, -0.5) {$\vdots$};
		\node [style=Aexp] (19) at (3.25, -0.5) {$\beta$};
		\node [style=none] (20) at (5.25, 0) {};
		\node [style=none] (22) at (4.25, -1) {};
		\node [style=none] (23) at (4.25, -0.5) {$\vdots$};
		\node [style=Aexp] (24) at (4.75, -0.5) {$\delta$};
		\node [style=Xexp] (25) at (2.5, 0) {};
		\node [style=Sdag] (30) at (0.25, 0) {};
		\node [style=Zexp] (31) at (1, 0) {};
		\node [style=Yexp] (32) at (4, 0) {};
	\end{pgfonlayer}
	\begin{pgfonlayer}{edgelayer}
		\draw (12.center) to (7.center);
		\draw (19.center) to (17.center);
		\draw (1.center) to (20.center);
		\draw (24.center) to (22.center);
		\draw (25) to (19);
		\draw (31) to (12);
		\draw (32) to (24);
	\end{pgfonlayer}
\end{tikzpicture}

%% file: figures/PauliPropsb.tikz
\begin{tikzpicture}[baseline={(1)}]
	\begin{pgfonlayer}{nodelayer}
		\node [style=none] (1) at (0.5, 0) {};
		\node [style=none] (7) at (1.25, -1) {};
		\node [style=none] (10) at (1.25, -0.5) {$\vdots$};
		\node [style=Aexp] (12) at (1.75, -0.5) {$\alpha$};
		\node [style=none] (17) at (2.75, -1) {};
		\node [style=none] (18) at (2.75, -0.5) {$\vdots$};
		\node [style=Aexp] (19) at (3.25, -0.5) {$\beta$};
		\node [style=none] (20) at (6, 0) {};
		\node [style=none] (22) at (4.25, -1) {};
		\node [style=none] (23) at (4.25, -0.5) {$\vdots$};
		\node [style=Aexp] (24) at (4.75, -0.5) {$-\delta$};
		\node [style=Sdag] (30) at (5.5, 0) {};
		\node [style=Zexp] (31) at (1, 0) {};
		\node [style=Yexp] (33) at (2.5, 0) {};
		\node [style=Xexp] (34) at (4, 0) {};
	\end{pgfonlayer}
	\begin{pgfonlayer}{edgelayer}
		\draw (12.center) to (7.center);
		\draw (19.center) to (17.center);
		\draw (1.center) to (20.center);
		\draw (24.center) to (22.center);
		\draw (31) to (12);
		\draw (33) to (19);
		\draw (34) to (24);
	\end{pgfonlayer}
\end{tikzpicture}

%% file: sections/related_work.tex
\label{sec:related_work}
\subsection{Trotter error}
Once a Pauli polynomial has been specified, an ordering of the Pauli gadgets needs to be determined. 
Normally, Pauli gadgets can only be reordered in a circuit if they commute, but since the gadgets emerge from a sum in the exponent (see \autoref{eq:trotter}), we can reorder the Pauli strings in the sum and, likewise, the corresponding Pauli gadgets, arbitrarily.

However, the transformation in \autoref{eq:trotter} is approximate and results in a small \emph{Trotter error}.
Although arbitrarily reordering the individual Pauli gadgets does not worsen the order of the approximation error in trotterized decompositions, investigations by \cite{Grimsley2019} indicate that the variance in Trotter error when reordering non-commuting terms is not trivial if interaction terms are large.

Investigations by \cite{Hastings2014}, \cite{Tranter2018}, \cite{cowtan2020generic}, and \cite{gui2021term} also indicate that the ordering of terms also impacts the number of entangling gates generated during synthesis, each proposing ordering strategies that reduce both synthesized gate count and theoretical upper bounds for the resultant Trotter error. Additionally, lexicographical ordering, in which the Pauli strings of each component are treated as words and sorted using a strict letter ordering, is shown to be a good heuristic for obtaining gate cancellations between synthesized Pauli gadgets~\cite{Tranter2018}, \cite{Hastings2014}.

In this work, similar to the assumptions made in \cite{li2022paulihedral} in their Paulihedral implementation, we ignore the Trotter error for arbitrarily ordering Pauli gadgets, focusing mainly on reducing entangling gate count. This choice is grounded in the assumption that entangling gates introduce greater error than that introduced by poor ordering. Furthermore, the effects of poor ordering can be mitigated by limiting time step size and utilizing higher order decomposition methods~\cite{low2019wellconditioned}. 

Nevertheless, the proposed algorithm can also be used to synthesize the Pauli gadgets in groups. Although, we do not use this feature in our experiments, nor do we provide a grouping heuristic.

\subsection{Related algorithms}\label{subsec:related_algorithms}
Once a grouping of Pauli gadgets is decided, various algorithms exist to synthesize these sets of gadgets. We use two of these as a baseline to the proposed algorithm.

First, we have \cite{li2022paulihedral}, which assumes a user-specified grouping and uses the coupling map to group the Pauli gadgets based on cumulative qubit distance and attempts to synthesize each layer such that the gates between layers are maximally canceled.

Secondly, we compare against \cite{cowtan2020generic} where Pauli gadgets are optionally re-grouped using graph coloring. Afterwards, each group is synthesized by diagonalizing the Pauli gadgets and then synthesizing them using GraySynth~\cite{Amy_2018}. The idea of \cite{Amy_2018} is to synthesize the gadgets in order of the Gray code~\cite{Gray1953}, such that each subsequent gadget only requires a single CNOT to be synthesized. 

In this paper, we propose an alternative method for ordering and synthesizing a sequence of Pauli gadgets that is inspired by GraySynth~\cite{Amy_2018} and its architecture-aware version~\cite{Meijer_van_de_Griend_2023}. 


\subsection{Clifford Tableaus}
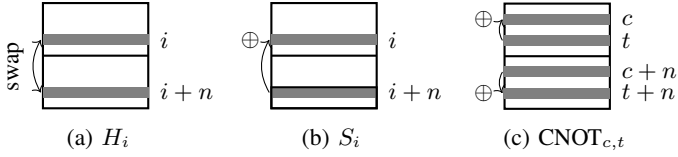
\begin{figure}
\centering
    \begin{subfigure}{0.3\linewidth}
        \centering
        \begin{tikzpicture}[scale=0.14]
            \fill[white] (0,0) rectangle (10,10);
            \draw[black, thick] (0,0) rectangle (10,10);
            \draw[black, thick] (0,5) -- (10,5);
    
            \fill[gray] (0,6) rectangle (10,7);
            \draw[<->, thin, bend left] (-0.2,1.5) to node[midway, sloped, above]{swap} (-0.2,6.5);
            \node[anchor=west] at (10.2,6.5) {$i$};
            
            \fill[gray] (0,1) rectangle (10,2);
    
            \node[anchor=west] at (10.2,1.5) {$i+n$};
        \end{tikzpicture}
        \caption{$H_i$}
    \end{subfigure}
    \hfill
    \begin{subfigure}{0.3\linewidth}
        \centering
        \begin{tikzpicture}[scale=0.14]
            \fill[white] (0,0) rectangle (10,10);
            \draw[black, thick] (0,0) rectangle (10,10);
            \draw[black, thick] (0,5) -- (10,5);
            \fill[gray] (0,6) rectangle (10,7);
            \draw[->, thin, bend left] (-0.2,1.5) to node[pos=1, left]{$\oplus$} (-0.2,6.5);
            \node[anchor=west] at (10.2,6.5) {$i$};
            
            \fill[gray] (0,1) rectangle (10,2);
            \draw (0,0) rectangle (10,2);
    
            \node[anchor=west] at (10.2,1.5) {$i+n$};
        \end{tikzpicture}
        \caption{$S_i$}
    \end{subfigure}
    \hfill
    \begin{subfigure}{0.3\linewidth}
        \centering
        \begin{tikzpicture}[scale=0.14]
            \fill[white] (0,0) rectangle (10,10);
            \draw[black, thick] (0,0) rectangle (10,10);
            \draw[black, thick] (0,5) -- (10,5);
            
            \fill[gray] (0,8) rectangle (10,9);
            \fill[gray] (0,6) rectangle (10,7);
            \draw[->, thin, bend left] (-0.2,6.5) to node[pos=1, left]{$\oplus$} (-0.2,8.5);
            \node[anchor=west] at (10.2,8.5) {$c$};
            \node[anchor=west] at (10.2,6.5) {$t$};
            
            \fill[gray] (0,3) rectangle (10,4);
            \fill[gray] (0,1) rectangle (10,2);
    
            \draw[<-, thin, bend left] (-0.2,1.5) to node[pos=0, left]{$\oplus$} (-0.2,3.5);
            \node[anchor=west] at (10.2,3.5) {$c+n$};
            \node[anchor=west] at (10.2,1.5) {$t+n$};
            
        \end{tikzpicture}
        \caption{$\text{CNOT}_{c, t}$}
    \end{subfigure}
\caption{Action of prepending Clifford gates $H$, $S$ and CNOT to an $n$ qubit  Clifford tableau.}
\label{fig:prepending_op}
\end{figure}

A final component required by our proposed algorithm is Clifford tableau synthesis. The Clifford tableau is a compact representation introduced by \cite{Aaronson_2004} that specifies a Clifford circuit or operator. More specifically, an $n$ qubit Clifford circuit can be designated by a $2n \times 2n$ matrix and a vector of size $2n$, all in $GF(2)$ and the tableau of an empty circuit is an identity matrix.
Appending and prepending Clifford gates to a Clifford operator corresponds to row and column operations on its Clifford tableau. The prepending operations are visualized in \autoref{fig:prepending_op}, and the appending operations act transposed with respect to prepending: they act on the respective columns, rather than the rows. 

Given a Clifford tableau, one can synthesize the equivalent quantum circuit as follows, starting with an empty circuit:
\begin{enumerate}
    \item Generate a Clifford gate and its adjoint.
    \item Append the gate to the circuit and prepend it to the tableau.
    \item Continue until the Clifford tableau is an identity matrix.
\end{enumerate}
In this paper, we adapted one of our previous algorithms~\cite{winderl2023architectureaware}, which uses the structure of the RowCol algorithm~\cite{Wu_2023} for CNOT synthesis and applies it to the Clifford tableau synthesis from \cite{berg2021simple}, \cite{Gidney_2021}.

%% file: sections/methods.tex
The proposed algorithm combines the ideas from the related algorithms into a single architecture-aware strategy for synthesizing these trotterized Hamiltonians. 
The algorithm expects a set of Pauli gadgets that can be decomposed in arbitrary order, and returns a quantum circuit implementing those Pauli gadgets.
The structure of this algorithm is an adaptation of GraySynth~\cite{Amy_2018} and its architecture-aware variant~\cite{Meijer_van_de_Griend_2023}. The outline of the algorithm is as follows, with the functions \texttt{diagonalize} and \texttt{disconnect} defined in \autoref{subsec:gate_placement}:
\begin{enumerate}
    \item\label{itm:idrecurse} Choose a qubit $q_{pivot}$ to remove from the problem, according to heuristic $h_{pivot}$.
    \item\label{itm:basecase} Synthesize any Pauli gadgets that are connected to only a single qubit: \texttt{diagonalize} them and place the $R_Z(\alpha)$ gate. Then, remove these gadgets from the problem.
    \item\label{itm:split} Split the Pauli gadgets into groups based on the letters on the legs on $q_{pivot}$.
    \item\label{itm:rec1} Take the Pauli gadgets that are not connected to $q_{pivot}$, i.e. with an $I$ on the leg, and go to step \ref{itm:idrecurse} with $q_{pivot}$ removed from the problem.
    \item\label{itm:neighbor} Choose another qubit $q_{neighbor}$ that is highly entangled with the chosen qubit according to the remaining Pauli gadgets, according to heuristic $h_{n}$.
    \item\label{itm:swap} If there are Pauli gadgets not connected to $q_{neighbor}$, \texttt{diagonalize} as many Pauli gadgets as possible with respect to the first qubit, and place two CNOT gates between the two qubits to \texttt{disconnect} the first qubit.
    \item\label{itm:disconnect} Otherwise, \texttt{disconnect} as many remaining Pauli gadgets as possible by placing at most two single qubit Clifford gates to \texttt{diagonalize} some Pauli gadgets (i.e. convert legs to $\{Y, Z\}$, see \autoref{subsec:gate_placement} for specifics) and a CNOT gate between the two chosen qubits to disconnect the gadgets (see \cite{cowtan2019phase} for an overview).
    \item\label{itm:split2} Repeat step \ref{itm:split}: split the remaining Pauli gadgets.
    \item\label{itm:rec2} Repeat step \ref{itm:rec1}: go to step \ref{itm:idrecurse} with disconnected Pauli gadgets and $q_{pivot}$ removed from the problem.
    \item \label{itm:rec3} Go to step \ref{itm:neighbor} with the remaining remaining Pauli gadgets.
\end{enumerate}
The adjoint of all synthesized Clifford gates are aggregated into a Clifford tableau and later synthesized (e.g. using \cite{winderl2023architectureaware}) and all unfinished Pauli gadgets are updated according to the Clifford gate commutation rules~\cite{cowtan2019phase} whenever a Clifford gate is synthesized.
Additionally, the algorithm can be adjusted to synthesize the Pauli gadgets in the given order by iteratively using this procedure on sets of mutually commuting Pauli gadgets. In which case, the Clifford tableau collects the Clifford gates from all iterations and the Clifford gates are propagated through all subsequent Pauli gadgets, including those in future iterations.
Moreover, this algorithm can be trivially made architecture-aware by choosing a $q_{pivot}$ in step \ref{itm:idrecurse} that does not disconnect the connectivity graph when removed, and restricting  $q_{neighbor}$  in step \ref{itm:neighbor} to neighbors of the chosen qubit. 

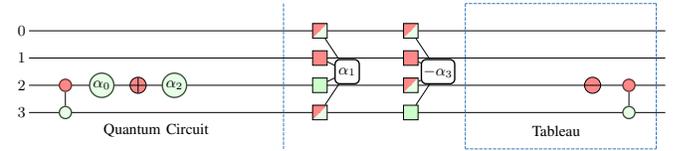
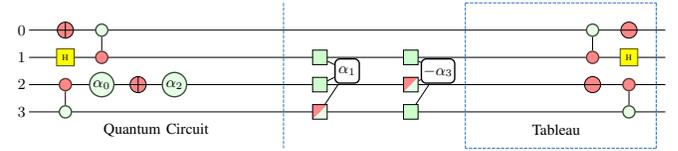
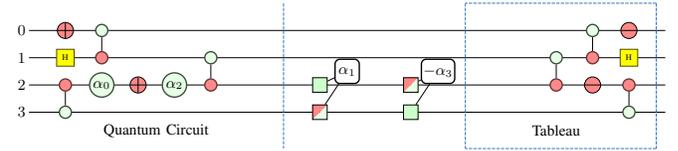
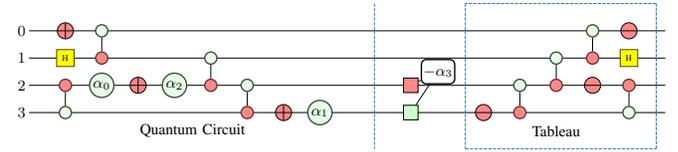
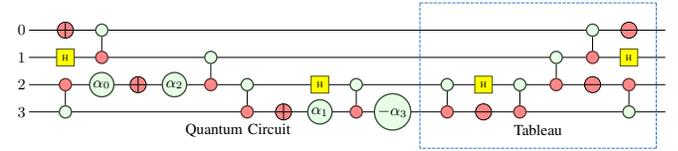
\begin{figure}[h!]
    \centering
    \begin{subfigure}{\linewidth}
    \resizebox{\linewidth}{!}{\input{\tikzitpathfigures/newexample_a.tikz}}
        \caption{Initial placement.}
    \end{subfigure}
    \par\bigskip
    \begin{subfigure}{\linewidth}
        \centering
        \resizebox{\linewidth}{!}{\input{\tikzitpathfigures/newexample_b.tikz}}
        \caption{Greedily group gadgets $\alpha_0$ and $\alpha_2$.}
    \end{subfigure}
    \par\bigskip
    \begin{subfigure}{\linewidth}
        \centering
        \resizebox{\linewidth}{!}{\input{\tikzitpathfigures/newexample_c.tikz}}
        \caption{\texttt{disconnect} gadgets $\alpha_0$ and $\alpha_2$, then \texttt{diagonalize}}
    \end{subfigure}
    \par\bigskip
    \begin{subfigure}{\linewidth}
        \centering
        \resizebox{\linewidth}{!}{\input{\tikzitpathfigures/newexample_d.tikz}}
        \caption{\texttt{diagonalize} qubit $0$ and \texttt{disconnect} it from gadgets $\alpha_1$ and $\alpha_3$.}
    \end{subfigure}
    \par\bigskip
    \begin{subfigure}{\linewidth}
        \centering
        \resizebox{\linewidth}{!}{\input{\tikzitpathfigures/newexample_e.tikz}}
        \caption{\texttt{disconnect} qubit $1$ from gadgets $\alpha_1$ and $\alpha_3$.}
    \end{subfigure}
    \par\bigskip
    \begin{subfigure}{\linewidth}
        \centering
        \resizebox{\linewidth}{!}{\input{\tikzitpathfigures/newexample_f.tikz}}
        \caption{\texttt{disconnect} qubit 3 to synthesize gadget $\alpha_1$.}
    \end{subfigure}
    \par\bigskip
    \begin{subfigure}{\linewidth}
        \centering
        \resizebox{\linewidth}{!}{\input{\tikzitpathfigures/newexample_g.tikz}}
        \caption{\texttt{disconnect} qubit 3 again to synthesize gadget $\alpha_3$.}
    \end{subfigure}
    \par\bigskip
   \caption{Step-by-step synthesis of time evolution operator $H=\alpha_0 IIZZ + \alpha_1 YXZX + \alpha_2 IIYZ + \alpha_3 YXZI$ from initial placement using methods in \autoref{subsec:gate_placement}.$^{\ref{note1}}$}
   \label{fig:alg_steps}
\end{figure}

Intuitively, this algorithm tries to greedily find an optimal lexicographical order for the letters, i.e. colors, on each Pauli gadget, per qubit. Because of the greedy nature, we assume that $I$ always comes first in the alphabet, such that we never unnecessarily introduce entanglement with a new qubit. The exception is when this is needed for routing purposes in the architecture-aware case. Moreover, as explained in the remainder of this section, when placing the Clifford gates, we essentially partially decompose the Pauli gadgets and assume that the gates on the other side of the Pauli gadget will cancel out. This is done by propagating the adjoint of the added Clifford gates through all remaining Pauli gadgets and gathering them in a Clifford tableau to be synthesized as a final optimization step. Thus, all Clifford gates are generated as late as possible, on an as-needed basis. As a result, the algorithm does not need to guess a good gadget ordering a priori. Instead, the ordering emerges from the structure of the synthesis procedure, while grouping gadgets with the same letters on the same qubits together. A simple example that steps through the synthesis process is shown in \autoref{fig:alg_steps}.

\subsection{Gate placement}
\label{subsec:gate_placement}
To be precise, we \texttt{diagonalize} gadgets with respect to a single qubit as follows: if the gadget has no leg or a green leg on the qubit (i.e. $I$, or $Z$), we are done, otherwise, we place a single qubit Clifford gate and its adjoint according to the following transformations:\footnote{\label{note1}Here, we use the dashed lines to separate the diagram into different regions: the left-most region contains the gates added to the circuit, the middle region is the updated Pauli gadgets, and the right-most region contains gates that are stored in a Clifford tableau for re-synthesis.}
\begin{equation}\label{eq:diagonalize_x}
\tag{\texttt{diagonalize} X}
    \input{\tikzitpathfigures/DiagonalizeXa.tikz} = \input{\tikzitpathfigures/DiagonalizeXb.tikz} = \input{\tikzitpathfigures/DiagonalizeXc.tikz}
\end{equation}\vspace{0.2pt}
\begin{equation}\label{eq:diagonalize_y}\tag{\texttt{diagonalize} Y}
    \input{\tikzitpathfigures/DiagonalizeYa.tikz} = \input{\tikzitpathfigures/DiagonalizeYb.tikz} = \input{\tikzitpathfigures/DiagonalizeYc.tikz}
\end{equation}\vspace{0.2pt}
The introduced gate is propagated through all following Pauli gadgets, and gathered in the Clifford tableau, following the rules in \autoref{fig:prepending_op}.
Note that we choose the rule that \texttt{diagonalize}s as many Pauli gadgets as possible. Typically, not all affected legs are mapped to a green leg and their updated type must be considered.

After diagonalizing, we \texttt{disconnect} as many Pauli gadgets as possible by placing up to a single qubit Clifford gate and two CNOTs. By construction, the first qubit will always be diagonalized.
The simplest case is when the second qubit is already disconnected. Then, we can swap the two qubits. However, because we know the type of the Pauli gadget, we can remove one of the CNOTs since it commutes with the Pauli gadget and cancels out. Thus, we have:$^{\ref{note1}}$
\begin{equation}\label{eq:swap}\tag{\texttt{disconnect} I}
    \input{\tikzitpathfigures/Swapa.tikz} = \input{\tikzitpathfigures/Swapb.tikz}
\end{equation}

If the second qubit is connected, we can match the legs of the Pauli gadgets under consideration as one of three cases, each corresponding to a different Clifford operation to be applied:$^{\ref{note1}}$
\begin{equation}\label{eq:disentangle_yz}\tag{\texttt{disconnect} Z or Y}
    \input{\tikzitpathfigures/DisentangleYZa.tikz} = \input{\tikzitpathfigures/DisentangleYZb.tikz}
\end{equation}\vspace{0.2pt}
\begin{equation}\label{eq:disentangle_xz}\tag{\texttt{disconnect} Z or X}
    \input{\tikzitpathfigures/DisentangleXZa.tikz} = \input{\tikzitpathfigures/DisentangleXZb.tikz}
\end{equation}\vspace{0.2pt}
\begin{equation}\label{eq:disentangle_xy}\tag{\texttt{disconnect} X or Y}
    \input{\tikzitpathfigures/DisentangleXYa.tikz} = \input{\tikzitpathfigures/DisentangleXYb.tikz}
\end{equation}

Each introduced rule only removes entanglement for two types of legs in the second wire. However, the second wire can contain $X$, $Y$ or $Z$ legs. Thus, we choose the option where the most Pauli gadgets are matched and place the corresponding Clifford gates. Then, we propagate these gates through all remaining Pauli gadgets and prepend them to the Clifford tableau.
\captionsetup[figure]{font=small, skip=0pt}

\subsection{Heuristics}
The proposed algorithm relies on choosing a good qubit to perform the splitting heuristic on. For our experiments, we have used the following heuristic and cost function:
\begin{align}
    c_{q}(s_q, \bf{g}) &= k \cdot P_{I}(s_q[\bf{g}]) \\ 
    &\phantom{{}=} +  P_{max}(s_q[\bf{g}])  -  P_{min}(s_q[\bf{g}]), \nonumber \\
    h(\bf{g}, \textbf{o}) &= \text{arg max}_{q \in \textbf{o}}(c_{q}(s_q, \bf{g})),  \nonumber \\
    h_{pivot}(\textbf{g}, G) &= h(\bf{g}, \{q \in \texttt{nodes}(G) \;| \; \neg\texttt{cuts}(q,G)\}),  \nonumber \\
    h_{n}(q_p, \textbf{g}, G) &= h(\bf{g}, \{q \in \texttt{nodes}(G) \;| \; (q_p, q)\in \texttt{edges}(G)\}),  \nonumber
\end{align}
where $q_p$ is the chosen pivot qubit, $s_q$ is the string of Pauli letters of all legs on qubit $q$, $\bf{g}$ the set of Pauli gadgets examined, and $k$ describes a weighting factor which we parametrized with $k=10$. $P_I(s)$, $P_{max}(s)$ and $P_{min}(s)$ are all functions acting on the Pauli string and returning a number in $\mathbf{N}$:
$P_I(s)$ counts the amount of  Pauli $I$'s in $s$, $P_{max}(s)$ counts the size of the largest non-$I$ region in $s$, $P_{min}(s)$ determines the size of the smallest non-$I$ region in $s$.

It is possible to tweak the cost function in our heuristic to be more specific to the target architecture. For example, the cost can be weighted based on the fidelity of the qubits or the gates, or adjusted to prefer gate parallelisation. In principle, any cost function would result in a correct algorithm, but with a different performance in terms of resulting gate count. We expect that the design of an optimal cost function is architecture dependent and leave analysis to future work.

\subsection{Clifford tableau synthesis up-to-permutation}
The last step of our algorithm is the synthesis of the Clifford tableau with the propagated Clifford gates. For this, we use an adaptation of the method proposed by \cite{winderl2023architectureaware}. We extend this algorithm to include the trick from \cite{MeijervandeGriend2023} that allows us to synthesize the tableau up-to-permutation; emulating the movement of qubits on the quantum computers. Thus, the resulting circuit might have qubits that are moved to a different register than they were originally placed in. This is a normal phenomenon in quantum circuits that have been routed using SWAP gates and it allows us to reduce CNOT count.

Similar to the proposed algorithm, the Clifford tableau synthesis algorithm~\cite{winderl2023architectureaware} uses a heuristic for picking which qubit to synthesize and remove from the problem. While the final Clifford tableau is typically an identity matrix, we can alternatively end the synthesis procedure when the Clifford tableau is a permutation matrix. The permutation then corresponds to a relocation of the qubits on the quantum computer. We do this by introducing a secondary heuristic for choosing which register (i.e. column in the tableau) the chosen qubit goes to, following the strategy from \cite{MeijervandeGriend2023}. Lastly, we need to take care that the signs of the qubits are corrected on the new qubit register and not the old one.

When selecting the logical qubit (row), we choose the following heuristic:
\begin{equation}
    \begin{aligned}
        h_{\text{row}}(r) & = \sum_i \left[\mathbb{I}(T_{r,i} \neq 0 \vee T_{r,i+q} \neq 0) \right.\\ &\left. \quad + \mathbb{I}(T_{r+q,i} \neq 0 \vee T_{r+q,i+q} \neq 0)\right],
    \end{aligned}
\end{equation}
where $T$ is a Clifford tableau, and $q$ is the number of qubits. This heuristic favors the row with the fewest interactions. When choosing a physical qubit (column), we look for the shortest CNOT distance, which is approximated by summing up the distances of rows along the connectivity graph:
\begin{equation}
    \begin{aligned}
        h_{\text{col}}(c) & = \sum_i \left[\mathbb{I}(T_{c,i} \neq 0 \vee T_{c,i+q} \neq 0) \right.\cdot d_{c, i}\\ &\left. \quad + \mathbb{I}(T_{c+q,i} \neq 0 \vee T_{c+q,i+q} \neq 0) \cdot d_{c, i}\right]
    \end{aligned}
\end{equation}
Here $d_{i,j}$, describes the shortest path between $i$ and $j$, something which can be efficiently precomputed using the Floyd-Warshall algorithm~\cite{floyd1962algorithm, warshall1962theorem}.

\subsection{Runtime Complexity Analysis}
The following recurrence relation describes the worst-case runtime of our algorithm:
\begin{equation}\label{eq:recurrance_relation}
	T(n) = \begin{cases}
		1 & n = 0 \\
		4T(n-1) + nq & \text{otherwise}
	\end{cases}
\end{equation}
where $n$ is the number of gadgets in the Pauli polynomial, and $q$ is the number of qubits. One can see that this is true since our method requires per step at most $\mathcal{O}(nq)$ many operations to compute the heuristics, reorder the gadgets, and propagate Clifford gates through the Pauli polynomial. Secondly, the worst case in splitting the Pauli polynomial is to select one gadget in one of the recursion steps and partition the rest to the others~\footnote{Note that this would be equivalent to decomposing one gadget and propagating the gates throughout the rest of the Pauli polynomial.}. Furthermore, we can see that \autoref{eq:recurrance_relation} is bound by $\mathcal{O}(n^2q + q^4)$, which yields the closed-form upper bound in terms of runtime. The term $\mathcal{O}(q^4)$ arises from synthesizing the Clifford tableau~\cite{winderl2023architectureaware}.

\subsection{CNOT Complexity Analysis}
Given that per $n$ of the gadgets, at most $q$ many CNOTs are required to synthesize it entirely, and given that there are at most $n$ gadgets, a bound can be found from the fact that there are at most $\mathcal{O}(nq)$ CNOTs required to synthesize the whole Pauli polynomial. Since the remaining gadgets are collected in a Clifford tableau, which requires at most $\mathcal{O}(q^2)$ many CNOTs among synthesis, we arrive at an overall overhead of $\mathcal{O}(nq + q^2)$ CNOTs. We want to emphasize that this bound is quite naive since duplicates are canceled when propagating the CNOTs through the rest of the Pauli Polynomial. So the average bound might be much tighter as indicated by our evaluations.

%% file: figures/DiagonalizeXa.tikz
\tikzset{OPLUS/.style={draw,circle,append after command={
        [shorten >=\pgflinewidth, shorten <=\pgflinewidth,]
        (\tikzlastnode.north) edge (\tikzlastnode.south)
        (\tikzlastnode.east) edge (\tikzlastnode.west)
        }
    }
}
\tikzset{OMINUS/.style={draw,circle,append after command={
        [shorten >=\pgflinewidth, shorten <=\pgflinewidth,]
        (\tikzlastnode.east) edge (\tikzlastnode.west)
        }
    }
}
\begin{tikzpicture}[baseline={(1)}]
	\begin{pgfonlayer}{nodelayer}
		
        \node [style=none] (wire_start) at (0.5, 0) {};
		\node [style=none] (wire_end) at (2.5, 0) {};
        \node [style=Xexp] (first_X) at (1.0, 0) {};
        \node [style=none] (otherleg_X) at (1.25, -1) {};
		
		\node [style=none] (alphadots) at (1.25, -0.5) {$\vdots$};
		\node [style=Aexp] (alpha) at (1.75, -0.5) {$\alpha$};
	\end{pgfonlayer}
	\begin{pgfonlayer}{edgelayer}
        \draw (wire_start) to (wire_end);
        \draw (first_X.center) to (alpha.center);
        \draw (otherleg_X.center) to (alpha.center);
	\end{pgfonlayer}
\end{tikzpicture}

%% file: figures/DiagonalizeXb.tikz
\begin{tikzpicture}[baseline={(1)}]
	\begin{pgfonlayer}{nodelayer}
        \node [style=none] (wire_start) at (-0.75, 0) {};
		\node [style=none] (wire_end) at (2.5, 0) {};
        \node [style=Xexp] (first_X) at (1.0, 0) {};
        \node [style=none] (otherleg_X) at (1.25, -1) {};
        \node [style=hadamard vertex] (hdag) at (-0.35, 0) {};
        \node [style=hadamard vertex] (h) at (0.25, 0) {};
		\node [style=none] (alphadots) at (1.25, -0.5) {$\vdots$};
		\node [style=Aexp] (alpha) at (1.75, -0.5) {$\alpha$};
	\end{pgfonlayer}
	\begin{pgfonlayer}{edgelayer}
        \draw (wire_start) to (wire_end);
        \draw (first_X.center) to (alpha.center);
        \draw (otherleg_X.center) to (alpha.center);
	\end{pgfonlayer}
\end{tikzpicture}

%% file: figures/DiagonalizeXc.tikz
\begin{tikzpicture}[baseline={(1)}]
	\begin{pgfonlayer}{nodelayer}
		
		\node [style=none] (wire_start) at (-0.5, 0) {};
		\node [style=none] (wire_end) at (3.2, 0) {};
        \node [style=Zexp] (first_Z) at (1.0, 0) {};
        \node [style=none] (otherleg_Z) at (1.25, -1) {};
        \node [style=hadamard vertex] (V) at (2.75, 0) {};		
        \node [style=hadamard vertex] (Vdag) at (0., 0) {};		
		
		\node [style=none] (alphadots) at (1.25, -0.5) {$\vdots$};
		\node [style=Aexp] (alpha) at (1.75, -0.5) {$\alpha$};
  
        \node [style=none, align=center, font=\tiny] (circuit) at (-0.15, -.65) {Quantum\\Circuit};
        \node [style=none, align=center, font=\tiny] (clifford) at (2.8, -.65) {Tableau};
	\end{pgfonlayer}
	\begin{pgfonlayer}{edgelayer}
        \draw (wire_start) to (wire_end);
        \draw (first_Z) to (alpha.center);
        \draw (otherleg_Z) to (alpha.center);
        \draw[dashed] (2.25, 1.) to (2.25, -1.);
        \draw[dashed] (0.5, 1.) to (0.5, -1.);
	\end{pgfonlayer}
\end{tikzpicture}

%% file: figures/DiagonalizeYa.tikz
\begin{tikzpicture}[baseline={(1)}]
	\begin{pgfonlayer}{nodelayer}
        \node [style=none] (wire_start) at (0.5, 0) {};
		\node [style=none] (wire_end) at (2.5, 0) {};
        \node [style=Yexp] (first_Y) at (1.0, 0) {};
        \node [style=none] (otherleg_Y) at (1.25, -1) {};
		
		\node [style=none] (alphadots) at (1.25, -0.5) {$\vdots$};
		\node [style=Aexp] (alpha) at (1.75, -0.5) {$\alpha$};
	\end{pgfonlayer}
	\begin{pgfonlayer}{edgelayer}
        \draw (wire_start) to (wire_end);
        \draw (first_Y.center) to (alpha.center);
        \draw (otherleg_Y.center) to (alpha.center);
	\end{pgfonlayer}
\end{tikzpicture}

%% file: figures/DiagonalizeYb.tikz
\begin{tikzpicture}[baseline={(1)}]
	\begin{pgfonlayer}{nodelayer}
		
        \node [style=none] (wire_start) at (-0.75, 0) {};
		\node [style=none] (wire_end) at (2.5, 0) {};
        \node [style=Yexp] (first_Y) at (1.0, 0) {};
        \node [style=none] (otherleg_Y) at (1.25, -1) {};
		
		\node [style=none] (alphadots) at (1.25, -0.5) {$\vdots$};
		\node [style=Aexp] (alpha) at (1.75, -0.5) {$\alpha$};

        \node [style=OPLUS, fill=red] (hdag) at (-0.35, 0) {};
        \node [style=OMINUS, fill=red] (h) at (0.25, 0) {};
	\end{pgfonlayer}
	\begin{pgfonlayer}{edgelayer}
        \draw (wire_start) to (wire_end);
        \draw (first_Y.center) to (alpha.center);
        \draw (otherleg_Y.center) to (alpha.center);
	\end{pgfonlayer}
\end{tikzpicture}

%% file: figures/DiagonalizeYc.tikz
\tikzset{OPLUS/.style={draw,circle,append after command={
        [shorten >=\pgflinewidth, shorten <=\pgflinewidth,]
        (\tikzlastnode.north) edge (\tikzlastnode.south)
        (\tikzlastnode.east) edge (\tikzlastnode.west)
        }
    }
}
\tikzset{OMINUS/.style={draw,circle,append after command={
        [shorten >=\pgflinewidth, shorten <=\pgflinewidth,]
        (\tikzlastnode.east) edge (\tikzlastnode.west)
        }
    }
}
\begin{tikzpicture}[baseline={(1)}]
	\begin{pgfonlayer}{nodelayer}
		
		\node [style=none] (wire_start) at (-0.5, 0) {};
		\node [style=none] (wire_end) at (3.2, 0) {};
        \node [style=Zexp] (first_Z) at (1.0, 0) {};
        \node [style=none] (otherleg_Z) at (1.25, -1) {};
        \node [style=OMINUS, fill=red] (V) at (2.75, 0) {};		
        \node [style=OPLUS, fill=red] (Vdag) at (0., 0) {};		
		
		\node [style=none] (alphadots) at (1.25, -0.5) {$\vdots$};
		\node [style=Aexp] (alpha) at (1.75, -0.5) {$\alpha$};

        \node [style=none, align=center, font=\tiny] (circuit) at (-0.15, -.65) {Quantum\\Circuit};
        \node [style=none, align=center, font=\tiny] (clifford) at (2.8, -.65) {Tableau};
	\end{pgfonlayer}
	\begin{pgfonlayer}{edgelayer}
        \draw (wire_start) to (wire_end);
        \draw (first_Z) to (alpha.center);
        \draw (otherleg_Z) to (alpha.center);
        
        \draw[dashed] (2.25, 1.) to (2.25, -1.);
        \draw[dashed] (0.5, 1.) to (0.5, -1.);
	\end{pgfonlayer}
\end{tikzpicture}

%% file: figures/Swapa.tikz
\begin{tikzpicture}[baseline={(1)}]
	\begin{pgfonlayer}{nodelayer}
		\node [style=none] (wire2_start) at (0.5, 0.5) {};
		\node [style=none] (wire1_start) at (0.5, 0) {};
        \node [style=none] (wire2_end) at (2.5, 0.5) {};
		\node [style=none] (wire1_end) at (2.5, 0) {};
  

		
		\node [style=none] (alpha_dots) at (1.25, -0.5) {$\vdots$};
        \node [style=Zexp] (alpha_z2) at (1., 0.5) {};
		\node [style=Aexp] (alpha) at (1.75, -0.5) {$\alpha$};
        \node [style=none] (alpha_lower) at (1.25, -1) {};


	\end{pgfonlayer}
	\begin{pgfonlayer}{edgelayer}
		\draw (wire1_start) to (wire1_end);
        \draw (wire2_start) to (wire2_end);
        \draw (alpha_z2) to (alpha);
        \draw (alpha_lower.center) to (alpha.center);


	\end{pgfonlayer}
\end{tikzpicture}

%% file: figures/Swapb.tikz
\begin{tikzpicture}[baseline={(1)}]
	\begin{pgfonlayer}{nodelayer}
		\node [style=none] (wire2_start) at (-1.2, 0.5) {};
		\node [style=none] (wire1_start) at (-1.2, 0) {};
        \node [style=none] (wire2_end) at (3.2, 0.5) {};
		\node [style=none] (wire1_end) at (3.2, 0) {};
  

        \node [style=red vertex] (cnot_x0) at (-0.75, 0.5) {};
        \node [style=green vertex] (cnot_z0) at (-0.75, 0) {};

		\node [style=green vertex] (cnot_z1) at (-0.25, 0.5) {};
        \node [style=red vertex] (cnot_x1) at (-0.25, 0) {};
		
		\node [style=none] (alpha_dots) at (0.75, -0.5) {$\vdots$};
		\node [style=Zexp] (alpha_z1) at (0.5, 0) {};
		\node [style=Aexp] (alpha) at (1.25, -0.5) {$\alpha$};
        \node [style=none] (alpha_lower) at (0.75, -1) {};


        \node [style=green vertex] (cnot_z2) at (2.25, 0.5) {};
        \node [style=red vertex] (cnot_x2) at (2.25, 0) {};

        \node [style=red vertex] (cnot_x3) at (2.75, 0.5) {};
        \node [style=green vertex] (cnot_z3) at (2.75, 0) {};
        \node [style=none, align=center, font=\tiny] (circuit) at (-0.6, -.65) {Quantum\\Circuit};
        \node [style=none, align=center, font=\tiny] (clifford) at (2.4, -.65) {Tableau};
	\end{pgfonlayer}
	\begin{pgfonlayer}{edgelayer}
		\draw (wire1_start) to (wire1_end);
        \draw (wire2_start) to (wire2_end);
        
        \draw(cnot_z0) to (cnot_x0);
        \draw(cnot_z1) to (cnot_x1);

        \draw (alpha_z1) to (alpha);
        \draw (alpha_lower.center) to (alpha.center);

        \draw[dashed] (1.75, 1.) to (1.75, -1.);
        \draw[dashed] (0.1, 1.) to (0.1, -1.);

        \draw(cnot_z2) to (cnot_x2);
        \draw(cnot_z3) to (cnot_x3);
	\end{pgfonlayer}
\end{tikzpicture}

%% file: figures/DisentangleYZa.tikz
\begin{tikzpicture}[baseline={(1)}]
	\begin{pgfonlayer}{nodelayer}
		\node [style=none] (wire2_start) at (0, 0.5) {};
		\node [style=none] (wire1_start) at (0, 0) {};
        \node [style=none] (wire2_end) at (3.25, 0.5) {};
		\node [style=none] (wire1_end) at (3.25, 0) {};
  

		\node [style=none] (alpha_dots) at (0.75, -0.5) {$\vdots$};
        \node [style=Zexp] (alpha_z2) at (0.5, 0.5) {};
		\node [style=Zexp] (alpha_z1) at (0.5, 0) {};
		\node [style=Aexp] (alpha) at (1.25, -0.5) {$\alpha$};
        \node [style=none] (alpha_lower) at (0.75, -1) {};

        \node [style=none] (beta_dpts) at (2.25, -0.5) {$\vdots$};
        \node [style=Zexp] (beta_z2) at (2., 0.5) {};
        \node [style=Yexp] (beta_y1) at (2., 0) {};
		\node [style=Aexp] (beta) at (2.75, -0.5) {$\beta$};
        \node [style=none] (beta_lower) at (2.25, -1) {};
	\end{pgfonlayer}
	\begin{pgfonlayer}{edgelayer}
		\draw (wire1_start) to (wire1_end);
        \draw (wire2_start) to (wire2_end);
        \draw (alpha_z2) to (alpha);
        \draw (alpha_z1) to (alpha);
        \draw (alpha_lower.center) to (alpha.center);

        \draw (beta_z2) to (beta);
        \draw (beta_y1) to (beta);
        \draw (beta_lower.center) to (beta.center);
        
	\end{pgfonlayer}
\end{tikzpicture}

%% file: figures/DisentangleYZb.tikz
\begin{tikzpicture}[baseline={(1)}]
	\begin{pgfonlayer}{nodelayer}
		\node [style=none] (wire2_start) at (-0.7, 0.5) {};
		\node [style=none] (wire1_start) at (-0.7, 0) {};
        \node [style=none] (wire2_end) at (4.2, 0.5) {};
		\node [style=none] (wire1_end) at (4.2, 0) {};
  

		\node [style=green vertex] (cnot_z1) at (-0.25, 0.5) {};
        \node [style=red vertex] (cnot_x1) at (-0.25, 0) {};
		
		\node [style=none] (alpha_dots) at (0.75, -0.5) {$\vdots$};
		\node [style=Zexp] (alpha_z1) at (0.5, 0) {};
		\node [style=Aexp] (alpha) at (1.25, -0.5) {$\alpha$};
        \node [style=none] (alpha_lower) at (0.75, -1) {};

        \node [style=none] (beta_dots) at (2.25, -0.5) {$\vdots$};
        \node [style=Yexp] (beta_y1) at (2., 0) {};
		\node [style=Aexp] (beta) at (2.75, -0.5) {$\beta$};
        \node [style=none] (beta_lower) at (2.25, -1) {};

        \node [style=green vertex] (cnot_z2) at (3.75, 0.5) {};
        \node [style=red vertex] (cnot_x2) at (3.75, 0) {};
        \node [style=none, align=center, font=\tiny] (circuit) at (-0.6, -.65) {Quantum\\Circuit};
        \node [style=none, align=center, font=\tiny] (clifford) at (3.8, -.65) {Tableau};
	\end{pgfonlayer}
	\begin{pgfonlayer}{edgelayer}
		\draw (wire1_start) to (wire1_end);
        \draw (wire2_start) to (wire2_end);
        \draw(cnot_z1) to (cnot_x1);
        \draw (alpha_z1) to (alpha);
        \draw (alpha_lower.center) to (alpha.center);

        \draw (beta_y1) to (beta);
        \draw (beta_lower.center) to (beta.center);
        \draw[dashed] (3.25, 1.) to (3.25, -1.);
        \draw[dashed] (0.1, 1.) to (0.1, -1.);
        \draw(cnot_z2) to (cnot_x2);
	\end{pgfonlayer}
\end{tikzpicture}

%% file: figures/DisentangleXZa.tikz
\begin{tikzpicture}[baseline={(1)}]
	\begin{pgfonlayer}{nodelayer}
		\node [style=none] (wire2_start) at (0, 0.5) {};
		\node [style=none] (wire1_start) at (0, 0) {};
        \node [style=none] (wire2_end) at (3.25, 0.5) {};
		\node [style=none] (wire1_end) at (3.25, 0) {};
  

		\node [style=none] (alpha_dots) at (0.75, -0.5) {$\vdots$};
        \node [style=Zexp] (alpha_z2) at (0.5, 0.5) {};
		\node [style=Zexp] (alpha_z1) at (0.5, 0) {};
		\node [style=Aexp] (alpha) at (1.25, -0.5) {$\alpha$};
        \node [style=none] (alpha_lower) at (0.75, -1) {};

        \node [style=none] (beta_dpts) at (2.25, -0.5) {$\vdots$};
        \node [style=Zexp] (beta_z2) at (2., 0.5) {};
        \node [style=Xexp] (beta_x1) at (2., 0) {};
		\node [style=Aexp] (beta) at (2.75, -0.5) {$\beta$};
        \node [style=none] (beta_lower) at (2.25, -1) {};
	\end{pgfonlayer}
	\begin{pgfonlayer}{edgelayer}
		\draw (wire1_start) to (wire1_end);
        \draw (wire2_start) to (wire2_end);
        \draw (alpha_z2) to (alpha.center);
        \draw (alpha_z1) to (alpha.center);
        \draw (alpha_lower.center) to (alpha.center);

        \draw (beta_z2) to (beta.center);
        \draw (beta_x1) to (beta.center);
        \draw (beta_lower.center) to (beta.center);
        
	\end{pgfonlayer}
\end{tikzpicture}

%% file: figures/DisentangleXZb.tikz
\begin{tikzpicture}[baseline={(1)}]
	\begin{pgfonlayer}{nodelayer}
		\node [style=none] (wire2_start) at (-1.2, 0.5) {};
		\node [style=none] (wire1_start) at (-1.2, 0) {};
        \node [style=none] (wire2_end) at (4.75, 0.5) {};
		\node [style=none] (wire1_end) at (4.75, 0) {};
  

		\node [style=green vertex] (cnot_z1) at (-0.25, 0.5) {};
        \node [style=red vertex] (cnot_x1) at (-0.25, 0) {};
		
		\node [style=none] (alpha_dots) at (0.75, -0.5) {$\vdots$};
		\node [style=Zexp] (alpha_z1) at (0.5, 0) {};
		\node [style=Aexp] (alpha) at (1.25, -0.5) {$\alpha$};
        \node [style=none] (alpha_lower) at (0.75, -1) {};

        \node [style=none] (beta_dots) at (2.25, -0.5) {$\vdots$};
        \node [style=Yexp] (beta_y1) at (2., 0) {};
		\node [style=Aexp] (beta) at (2.75, -0.5) {$-\beta$};
        \node [style=none] (beta_lower) at (2.25, -1) {};

        \node [style=green vertex] (cnot_z2) at (3.75, 0.5) {};
        \node [style=red vertex] (cnot_x2) at (3.75, 0) {};

        \node [style=OMINUS, fill=green] (Vdag) at (-0.75, 0) {};
        \node [style=OPLUS, fill=green] (V) at (4.25, 0) {};
        \node [style=none, align=center, font=\tiny] (circuit) at (-0.6, -.65) {Quantum\\Circuit};
        \node [style=none, align=center, font=\tiny] (clifford) at (4.0, -.65) {Tableau};
	\end{pgfonlayer}
	\begin{pgfonlayer}{edgelayer}
		\draw (wire1_start) to (wire1_end);
        \draw (wire2_start) to (wire2_end);
        \draw(cnot_z1) to (cnot_x1);
        \draw (alpha_z1) to (alpha);
        \draw (alpha_lower.center) to (alpha.center);

        \draw (beta_y1) to (beta);
        \draw (beta_lower.center) to (beta.center);

        \draw(cnot_z2) to (cnot_x2);
        \draw[dashed] (3.25, 1.) to (3.25, -1.);
        \draw[dashed] (0.1, 1.) to (0.1, -1.);
	\end{pgfonlayer}
\end{tikzpicture}

%% file: figures/DisentangleXYa.tikz
\begin{tikzpicture}[baseline={(1)}]
	\begin{pgfonlayer}{nodelayer}
		\node [style=none] (wire2_start) at (0, 0.5) {};
		\node [style=none] (wire1_start) at (0, 0) {};
        \node [style=none] (wire2_end) at (3.25, 0.5) {};
		\node [style=none] (wire1_end) at (3.25, 0) {};
  

		\node [style=none] (alpha_dots) at (0.75, -0.5) {$\vdots$};
        \node [style=Zexp] (alpha_z2) at (0.5, 0.5) {};
		\node [style=Xexp] (alpha_z1) at (0.5, 0) {};
		\node [style=Aexp] (alpha) at (1.25, -0.5) {$\alpha$};
        \node [style=none] (alpha_lower) at (0.75, -1) {};

        \node [style=none] (beta_dpts) at (2.25, -0.5) {$\vdots$};
        \node [style=Zexp] (beta_z2) at (2., 0.5) {};
        \node [style=Yexp] (beta_x1) at (2., 0) {};
		\node [style=Aexp] (beta) at (2.75, -0.5) {$\beta$};
        \node [style=none] (beta_lower) at (2.25, -1) {};
	\end{pgfonlayer}
	\begin{pgfonlayer}{edgelayer}
		\draw (wire1_start) to (wire1_end);
        \draw (wire2_start) to (wire2_end);
        \draw (alpha_z2) to (alpha.center);
        \draw (alpha_z1) to (alpha.center);
        \draw (alpha_lower.center) to (alpha.center);

        \draw (beta_z2) to (beta.center);
        \draw (beta_x1) to (beta.center);
        \draw (beta_lower.center) to (beta.center);

	\end{pgfonlayer}
\end{tikzpicture}

%% file: figures/DisentangleXYb.tikz
\begin{tikzpicture}[baseline={(1)}]
	\begin{pgfonlayer}{nodelayer}
		\node [style=none] (wire2_start) at (-1.2, 0.5) {};
		\node [style=none] (wire1_start) at (-1.2, 0) {};
        \node [style=none] (wire2_end) at (4.75, 0.5) {};
		\node [style=none] (wire1_end) at (4.75, 0) {};
  

		\node [style=green vertex] (cnot_z1) at (-0.25, 0.5) {};
        \node [style=red vertex] (cnot_x1) at (-0.25, 0) {};
		
		\node [style=none] (alpha_dots) at (0.75, -0.5) {$\vdots$};
		\node [style=Zexp] (alpha_z1) at (0.5, 0) {};
		\node [style=Aexp] (alpha) at (1.25, -0.5) {$\alpha$};
        \node [style=none] (alpha_lower) at (0.75, -1) {};

        \node [style=none] (beta_dots) at (2.25, -0.5) {$\vdots$};
        \node [style=Yexp] (beta_y1) at (2., 0) {};
		\node [style=Aexp] (beta) at (2.75, -0.5) {$\beta$};
        \node [style=none] (beta_lower) at (2.25, -1) {};

        \node [style=green vertex] (cnot_z2) at (3.75, 0.5) {};
        \node [style=red vertex] (cnot_x2) at (3.75, 0) {};

        \node [style=hadamard vertex] (Hdag) at (-0.75, 0) {};
        \node [style=hadamard vertex] (H) at (4.25, 0) {};
        \node [style=none, align=center, font=\tiny] (circuit) at (-0.6, -.65) {Quantum\\Circuit};
        \node [style=none, align=center, font=\tiny] (clifford) at (4.0, -.65) {Tableau};
	\end{pgfonlayer}
	\begin{pgfonlayer}{edgelayer}
		\draw (wire1_start) to (wire1_end);
        \draw (wire2_start) to (wire2_end);
        \draw(cnot_z1) to (cnot_x1);
        \draw (alpha_z1) to (alpha);
        \draw (alpha_lower.center) to (alpha.center);

        \draw (beta_y1) to (beta);
        \draw (beta_lower.center) to (beta.center);

        \draw(cnot_z2) to (cnot_x2);
        \draw[dashed] (3.25, 1.) to (3.25, -1.);
        \draw[dashed] (0.1, 1.) to (0.1, -1.);
	\end{pgfonlayer}
\end{tikzpicture}

%% file: figures/newexample_a.tikz
\begin{tikzpicture}[baseline={(13)}]
	\begin{pgfonlayer}{nodelayer}
        \node [style=none] (q0) at (-2.2, 0) {$3$};
		\node [style=none] (q1) at (-2.2, 0.75) {$2$};
		\node [style=none] (q2) at (-2.2, 1.5) {$1$};
		\node [style=none] (q3) at (-2.2, 2.25) {$0$};
		\node [style=none] (0) at (-2, 0) {};
		\node [style=none] (1) at (-2, 0.75) {};
		\node [style=none] (2) at (-2, 1.5) {};
		\node [style=none] (3) at (-2, 2.25) {};
		\node [style=none] (9) at (15.5, 2.25) {};
		\node [style=none] (10) at (15.5, 1.5) {};
		\node [style=none] (11) at (15.5, 0.75) {};
		\node [style=none] (12) at (15.5, 0) {};
		\node [style=Aexp] (13) at (1.75, 1.125) {$\alpha_0$};
		\node [style=none] (14) at (-2, 0) {};
		\node [style=none] (15) at (2.5, 0.75) {};
		\node [style=none] (16) at (-2, 1.5) {};
		\node [style=none] (17) at (2.5, 2.25) {};
		\node [style=none] (21) at (5, 2.25) {};
		\node [style=none] (22) at (5, 1.5) {};
		\node [style=none] (23) at (5, 0.75) {};
		\node [style=none] (24) at (5, 0) {};
		\node [style=Aexp] (25) at (4.25, 1.125) {$\alpha_1$};
		\node [style=none] (26) at (5, 0) {};
		\node [style=none] (27) at (5, 0.75) {};
		\node [style=none] (28) at (5, 1.5) {};
		\node [style=none] (29) at (5, 2.25) {};
		\node [style=none] (33) at (7.5, 2.25) {};
		\node [style=none] (34) at (7.5, 1.5) {};
		\node [style=none] (35) at (7.5, 0.75) {};
		\node [style=none] (36) at (7.5, 0) {};
		\node [style=Aexp] (37) at (6.75, 1.125) {$\alpha_2$};
		\node [style=none] (38) at (7.5, 0) {};
		\node [style=none] (39) at (7.5, 0.75) {};
		\node [style=none] (40) at (15.5, 1.5) {};
		\node [style=none] (41) at (7.5, 2.25) {};
		\node [style=none] (44) at (15.5, 2.25) {};
		\node [style=none] (46) at (15.5, 0.75) {};
		\node [style=none] (47) at (15.5, 0) {};
		\node [style=Aexp] (48) at (9.25, 1.125) {$\alpha_3$};
		\node [style=none] (49) at (0, 3) {};
		\node [style=none] (50) at (0, -1) {};
		\node [style=Zexp] (53) at (1, 0.75) {};
		\node [style=Zexp] (54) at (1, 0) {};
		\node [style=Yexp] (59) at (6, 0.75) {};
		\node [style=Zexp] (60) at (6, 0) {};
		\node [style=Xexp] (62) at (8.5, 1.5) {};
		\node [style=Yexp] (66) at (3.5, 2.25) {};
		\node [style=Xexp] (67) at (3.5, 1.5) {};
		\node [style=Zexp] (68) at (3.5, 0.75) {};
		\node [style=Yexp] (69) at (8.5, 2.25) {};
		\node [style=Xexp] (70) at (3.5, 0) {};
		\node [style=Zexp] (71) at (8.5, 0.75) {};
		\node [style=none] (72) at (10, 3) {};
		\node [style=none] (73) at (10, -1) {};
		\node [style=none] (74) at (15.25, -1) {};
		\node [style=none] (75) at (15.25, 3) {};
		\node [style=none] (77) at (12.5, -0.5) {Tableau};
		\node [style=none, align=center] (79) at (-1.05, -0.5) {Quantum\\Circuit};
	\end{pgfonlayer}
	\begin{pgfonlayer}{edgelayer}
		\draw (3.center) to (9.center);
        \draw (2.center) to (10.center);
        \draw (1.center) to (11.center);
        \draw (0.center) to (12.center);
		\draw [style=hadamard edge] (49.center) to (50.center);
		\draw (53) to (13);
		\draw (54) to (13);
		\draw (59) to (37);
		\draw (60) to (37);
		\draw (62) to (48);
		\draw (66) to (25);
		\draw (67) to (25);
		\draw (68) to (25);
		\draw (69) to (48);
		\draw (25) to (70);
		\draw (71) to (48);
		\draw [style=hadamard edge] (72.center) to (75.center);
		\draw [style=hadamard edge] (75.center) to (74.center);
		\draw [style=hadamard edge] (74.center) to (73.center);
		\draw [style=hadamard edge] (72.center) to (73.center);
	\end{pgfonlayer}
\end{tikzpicture}

%% file: figures/newexample_b.tikz
\begin{tikzpicture}[baseline={(13)}]
	\begin{pgfonlayer}{nodelayer}
         \node [style=none] (q0) at (-2.2, 0) {$3$};
		\node [style=none] (q1) at (-2.2, 0.75) {$2$};
		\node [style=none] (q2) at (-2.2, 1.5) {$1$};
		\node [style=none] (q3) at (-2.2, 2.25) {$0$};
		\node [style=none] (0) at (-2, 0) {};
		\node [style=none] (1) at (-2, 0.75) {};
		\node [style=none] (2) at (-2, 1.5) {};
		\node [style=none] (3) at (-2, 2.25) {};
		\node [style=none] (9) at (15.5, 2.25) {};
		\node [style=none] (10) at (15.5, 1.5) {};
		\node [style=none] (11) at (15.5, 0.75) {};
		\node [style=none] (12) at (15.5, 0) {};
		\node [style=Aexp] (13) at (1.75, 1.125) {$\alpha_0$};
		\node [style=none] (14) at (-2, 0) {};
		\node [style=none] (15) at (2.5, 0.75) {};
		\node [style=none] (16) at (-2, 1.5) {};
		\node [style=none] (17) at (2.5, 2.25) {};
		\node [style=none] (21) at (5, 2.25) {};
		\node [style=none] (22) at (5, 1.5) {};
		\node [style=none] (23) at (5, 0.75) {};
		\node [style=none] (24) at (5, 0) {};
		\node [style=Aexp] (25) at (4.25, 1.125) {$\alpha_2$};
		\node [style=none] (26) at (5, 0) {};
		\node [style=none] (27) at (5, 0.75) {};
		\node [style=none] (28) at (5, 1.5) {};
		\node [style=none] (29) at (5, 2.25) {};
		\node [style=none] (33) at (7.5, 2.25) {};
		\node [style=none] (34) at (7.5, 1.5) {};
		\node [style=none] (35) at (7.5, 0.75) {};
		\node [style=none] (36) at (7.5, 0) {};
		\node [style=Aexp] (37) at (6.75, 1.125) {$\alpha_1$};
		\node [style=none] (38) at (7.5, 0) {};
		\node [style=none] (39) at (7.5, 0.75) {};
		\node [style=none] (40) at (15.5, 1.5) {};
		\node [style=none] (41) at (7.5, 2.25) {};
		\node [style=none] (44) at (15.5, 2.25) {};
		\node [style=none] (46) at (15.5, 0.75) {};
		\node [style=none] (47) at (15.5, 0) {};
		\node [style=Aexp] (48) at (9.25, 1.125) {$\alpha_3$};
		\node [style=none] (49) at (0, 3) {};
		\node [style=none] (50) at (0, -1) {};
		\node [style=none] (51) at (10, 3) {};
		\node [style=none] (52) at (10, -1) {};
		\node [style=Zexp] (53) at (1, 0.75) {};
		\node [style=Zexp] (54) at (1, 0) {};
		\node [style=Xexp] (62) at (8.5, 1.5) {};
		\node [style=Yexp] (69) at (8.5, 2.25) {};
		\node [style=Yexp] (70) at (3.5, 0.75) {};
		\node [style=Zexp] (71) at (3.5, 0) {};
		\node [style=Yexp] (72) at (6, 2.25) {};
		\node [style=Xexp] (73) at (6, 1.5) {};
		\node [style=Zexp] (74) at (6, 0.75) {};
		\node [style=Xexp] (75) at (6, 0) {};
		\node [style=Zexp] (76) at (8.5, 0.75) {};
		
		\node [style=none] (78) at (15.25, 3) {};
		\node [style=none] (79) at (15.25, -1) {};
        \node [style=none] (77) at (12.5, -0.5) {Tableau};
        \node [style=none, align=center] (80) at (-1.05, -0.5) {Quantum\\Circuit};
	\end{pgfonlayer}
	\begin{pgfonlayer}{edgelayer}
 		\draw (3.center) to (9.center);
        \draw (2.center) to (10.center);
        \draw (1.center) to (11.center);
        \draw (0.center) to (12.center);
		\draw [style=hadamard edge] (49.center) to (50.center);
		\draw [style=hadamard edge] (51.center) to (52.center);
		\draw (53) to (13);
		\draw (54) to (13);
		\draw (62) to (48);
		\draw (69) to (48);
		\draw (70) to (25);
		\draw (71) to (25);
		\draw (72) to (37);
		\draw (74) to (37);
		\draw (73) to (37);
		\draw (75) to (37);
		\draw (76) to (48);
		\draw [style=hadamard edge] (51.center) to (78.center);
		\draw [style=hadamard edge] (78.center) to (79.center);
		\draw [style=hadamard edge] (79.center) to (52.center);
	\end{pgfonlayer}
\end{tikzpicture}

%% file: figures/newexample_c.tikz
\begin{tikzpicture}[baseline={(13)}]
	\begin{pgfonlayer}{nodelayer}
        \node [style=none] (q0) at (-2.2, 0) {$3$};
		\node [style=none] (q1) at (-2.2, 0.75) {$2$};
		\node [style=none] (q2) at (-2.2, 1.5) {$1$};
		\node [style=none] (q3) at (-2.2, 2.25) {$0$};
		\node [style=none] (0) at (-2, 0) {};
		\node [style=none] (1) at (-2, 0.75) {};
		\node [style=none] (2) at (-2, 1.5) {};
		\node [style=none] (3) at (-2, 2.25) {};
		\node [style=none] (9) at (15.5, 2.25) {};
		\node [style=none] (10) at (15.5, 1.5) {};
		\node [style=none] (11) at (15.5, 0.75) {};
		\node [style=none] (12) at (15.5, 0) {};
		\node [style=none] (14) at (-2, 0) {};
		\node [style=none] (15) at (2.5, 0.75) {};
		\node [style=none] (16) at (-2, 1.5) {};
		\node [style=none] (17) at (2.5, 2.25) {};
		\node [style=none] (21) at (5, 2.25) {};
		\node [style=none] (22) at (5, 1.5) {};
		\node [style=none] (23) at (5, 0.75) {};
		\node [style=none] (24) at (5, 0) {};
		\node [style=none] (26) at (5, 0) {};
		\node [style=none] (27) at (5, 0.75) {};
		\node [style=none] (28) at (5, 1.5) {};
		\node [style=none] (29) at (5, 2.25) {};
		\node [style=none] (33) at (7.5, 2.25) {};
		\node [style=none] (34) at (7.5, 1.5) {};
		\node [style=none] (35) at (7.5, 0.75) {};
		\node [style=none] (36) at (7.5, 0) {};
		\node [style=Aexp] (37) at (6.75, 1.125) {$\alpha_1$};
		\node [style=none] (38) at (7.5, 0) {};
		\node [style=none] (39) at (7.5, 0.75) {};
		\node [style=none] (40) at (15.5, 1.5) {};
		\node [style=none] (41) at (7.5, 2.25) {};
		\node [style=none] (44) at (15.5, 2.25) {};
		\node [style=none] (46) at (15.5, 0.75) {};
		\node [style=none] (47) at (15.5, 0) {};
		\node [style=Aexp] (48) at (9.25, 1.125) {$-\alpha_3$};
		\node [style=none] (49) at (5, 3) {};
		\node [style=none] (50) at (5, -1) {};
		\node [style=none] (51) at (10, 3) {};
		\node [style=none] (52) at (10, -1) {};
		\node [style=Xexp] (62) at (8.5, 1.5) {};
		\node [style=Zexp] (63) at (8.5, 0) {};
		\node [style=Yexp] (69) at (8.5, 2.25) {};
		\node [style=Yexp] (72) at (6, 2.25) {};
		\node [style=Xexp] (73) at (6, 1.5) {};
		\node [style=green vertex] (77) at (-1, 0) {};
		\node [style=red vertex] (78) at (-1, 0.75) {};
		\node [style=green vertex] (79) at (0, 0.75) {$\alpha_0$};
		\node [style=green vertex] (81) at (2, 0.75) {$\alpha_2$};
		\node [style=green vertex] (82) at (14.5, 0) {};
		\node [style=red vertex] (83) at (14.5, 0.75) {};
		\node [style=Yexp] (85) at (8.5, 0.75) {};
		\node [style=Zexp] (88) at (6, 0.75) {};
		\node [style=Yexp] (89) at (6, 0) {};
		\node [style=Vdag] (90) at (13.5, 0.75) {};
		\node [style=V] (91) at (1, 0.75) {};
		\node [style=none] (92) at (1.5, -0.5) {Quantum Circuit};
		\node [style=none] (93) at (15.25, 3) {};
		\node [style=none] (94) at (15.25, -1) {};
		\node [style=none] (95) at (12.5, -0.5) {Tableau};
	\end{pgfonlayer}
	\begin{pgfonlayer}{edgelayer}
  		\draw (3.center) to (9.center);
        \draw (2.center) to (10.center);
        \draw (1.center) to (11.center);
        \draw (0.center) to (12.center);
		\draw [style=hadamard edge] (49.center) to (50.center);
		\draw [style=hadamard edge] (51.center) to (52.center);
		\draw (62) to (48);
		\draw (63) to (48);
		\draw (69) to (48);
		\draw (72) to (37);
		\draw (73) to (37);
		\draw (83) to (82);
		\draw (78) to (77);
		\draw (85) to (48);
		\draw (88) to (37);
		\draw (89) to (37);
		\draw [style=hadamard edge] (52.center) to (94.center);
		\draw [style=hadamard edge] (94.center) to (93.center);
		\draw [style=hadamard edge] (93.center) to (51.center);
	\end{pgfonlayer}
\end{tikzpicture}

%% file: figures/newexample_d.tikz
\begin{tikzpicture}[baseline={(13)}]
	\begin{pgfonlayer}{nodelayer}
        \node [style=none] (q0) at (-2.2, 0) {$3$};
		\node [style=none] (q1) at (-2.2, 0.75) {$2$};
		\node [style=none] (q2) at (-2.2, 1.5) {$1$};
		\node [style=none] (q3) at (-2.2, 2.25) {$0$};
		\node [style=none] (0) at (-2, 0) {};
		\node [style=none] (1) at (-2, 0.75) {};
		\node [style=none] (2) at (-2, 1.5) {};
		\node [style=none] (3) at (-2, 2.25) {};
		\node [style=none] (9) at (15.5, 2.25) {};
		\node [style=none] (10) at (15.5, 1.5) {};
		\node [style=none] (11) at (15.5, 0.75) {};
		\node [style=none] (12) at (15.5, 0) {};
		\node [style=none] (14) at (-2, 0) {};
		\node [style=none] (15) at (2.5, 0.75) {};
		\node [style=none] (16) at (-2, 1.5) {};
		\node [style=none] (17) at (2.5, 2.25) {};
		\node [style=none] (21) at (5, 2.25) {};
		\node [style=none] (22) at (5, 1.5) {};
		\node [style=none] (23) at (5, 0.75) {};
		\node [style=none] (24) at (5, 0) {};
		\node [style=none] (26) at (5, 0) {};
		\node [style=none] (27) at (5, 0.75) {};
		\node [style=none] (28) at (5, 1.5) {};
		\node [style=none] (29) at (5, 2.25) {};
		\node [style=none] (33) at (7.5, 2.25) {};
		\node [style=none] (34) at (7.5, 1.5) {};
		\node [style=none] (35) at (7.5, 0.75) {};
		\node [style=none] (36) at (7.5, 0) {};
		\node [style=Aexp] (37) at (6.75, 1.125) {$\alpha_1$};
		\node [style=none] (38) at (7.5, 0) {};
		\node [style=none] (39) at (7.5, 0.75) {};
		\node [style=none] (40) at (15.5, 1.5) {};
		\node [style=none] (41) at (7.5, 2.25) {};
		\node [style=none] (44) at (15.5, 2.25) {};
		\node [style=none] (46) at (15.5, 0.75) {};
		\node [style=none] (47) at (15.5, 0) {};
		\node [style=Aexp] (48) at (9.25, 1.125) {$-\alpha_3$};
		\node [style=none] (49) at (5, 3) {};
		\node [style=none] (50) at (5, -1) {};
		\node [style=none] (51) at (10, 3) {};
		\node [style=none] (52) at (10, -1) {};
		\node [style=Zexp] (63) at (8.5, 0) {};
		\node [style=green vertex] (77) at (-1, 0) {};
		\node [style=red vertex] (78) at (-1, 0.75) {};
		\node [style=green vertex] (79) at (0, 0.75) {$\alpha_0$};
		\node [style=green vertex] (81) at (2, 0.75) {$\alpha_2$};
		\node [style=green vertex] (82) at (14.5, 0) {};
		\node [style=red vertex] (83) at (14.5, 0.75) {};
		\node [style=Yexp] (85) at (8.5, 0.75) {};
		\node [style=Zexp] (88) at (6, 0.75) {};
		\node [style=Yexp] (89) at (6, 0) {};
		\node [style=hadamard vertex] (90) at (-1, 1.5) {};
		\node [style=hadamard vertex] (91) at (14.5, 1.5) {};
		\node [style=V] (92) at (-1, 2.25) {};
		\node [style=Vdag] (93) at (14.5, 2.25) {};
		\node [style=green vertex] (94) at (0, 2.25) {};
		\node [style=red vertex] (95) at (0, 1.5) {};
		\node [style=red vertex] (96) at (13.5, 1.5) {};
		\node [style=green vertex] (97) at (13.5, 2.25) {};
		\node [style=Zexp] (98) at (6, 1.5) {};
		\node [style=Zexp] (99) at (8.5, 1.5) {};
		\node [style=V] (100) at (1, 0.75) {};
		\node [style=Vdag] (101) at (13.5, 0.75) {};
		\node [style=none] (102) at (1.5, -0.5) {Quantum Circuit};
        \node [style=none] (103) at (12.5, -0.5) {Tableau};
		\node [style=none] (104) at (15.25, 3) {};
		\node [style=none] (105) at (15.25, -1) {};
	\end{pgfonlayer}
	\begin{pgfonlayer}{edgelayer}
  		\draw (3.center) to (9.center);
        \draw (2.center) to (10.center);
        \draw (1.center) to (11.center);
        \draw (0.center) to (12.center);
		\draw [style=hadamard edge] (49.center) to (50.center);
		\draw [style=hadamard edge] (51.center) to (52.center);
		\draw (63) to (48);
		\draw (83) to (82);
		\draw (78) to (77);
		\draw (85) to (48);
		\draw (88) to (37);
		\draw (89) to (37);
		\draw (94) to (95);
		\draw (97) to (96);
		\draw (98) to (37);
		\draw (99) to (48);
		\draw [style=hadamard edge] (51.center) to (104.center);
		\draw [style=hadamard edge] (104.center) to (105.center);
		\draw [style=hadamard edge] (105.center) to (52.center);
	\end{pgfonlayer}
\end{tikzpicture}

%% file: figures/newexample_e.tikz
\begin{tikzpicture}[baseline={(13)}]
	\begin{pgfonlayer}{nodelayer}
        \node [style=none] (q0) at (-2.2, 0) {$3$};
		\node [style=none] (q1) at (-2.2, 0.75) {$2$};
		\node [style=none] (q2) at (-2.2, 1.5) {$1$};
		\node [style=none] (q3) at (-2.2, 2.25) {$0$};
		\node [style=none] (0) at (-2, 0) {};
		\node [style=none] (1) at (-2, 0.75) {};
		\node [style=none] (2) at (-2, 1.5) {};
		\node [style=none] (3) at (-2, 2.25) {};
		\node [style=none] (9) at (15.5, 2.25) {};
		\node [style=none] (10) at (15.5, 1.5) {};
		\node [style=none] (11) at (15.5, 0.75) {};
		\node [style=none] (12) at (15.5, 0) {};
		\node [style=none] (14) at (-2, 0) {};
		\node [style=none] (15) at (2.5, 0.75) {};
		\node [style=none] (16) at (-2, 1.5) {};
		\node [style=none] (17) at (2.5, 2.25) {};
		\node [style=none] (21) at (5, 2.25) {};
		\node [style=none] (22) at (5, 1.5) {};
		\node [style=none] (23) at (5, 0.75) {};
		\node [style=none] (24) at (5, 0) {};
		\node [style=none] (26) at (5, 0) {};
		\node [style=none] (27) at (5, 0.75) {};
		\node [style=none] (28) at (5, 1.5) {};
		\node [style=none] (29) at (5, 2.25) {};
		\node [style=none] (33) at (7.5, 2.25) {};
		\node [style=none] (34) at (7.5, 1.5) {};
		\node [style=none] (35) at (7.5, 0.75) {};
		\node [style=none] (36) at (7.5, 0) {};
		\node [style=Aexp] (37) at (6.75, 1.125) {$\alpha_1$};
		\node [style=none] (38) at (7.5, 0) {};
		\node [style=none] (39) at (7.5, 0.75) {};
		\node [style=none] (40) at (15.5, 1.5) {};
		\node [style=none] (41) at (7.5, 2.25) {};
		\node [style=none] (44) at (15.5, 2.25) {};
		\node [style=none] (46) at (15.5, 0.75) {};
		\node [style=none] (47) at (15.5, 0) {};
		\node [style=Aexp] (48) at (9.25, 1.125) {$-\alpha_3$};
		\node [style=none] (49) at (5, 3) {};
		\node [style=none] (50) at (5, -1) {};
		\node [style=none] (51) at (10, 3) {};
		\node [style=none] (52) at (10, -1) {};
		\node [style=Zexp] (63) at (8.5, 0) {};
		\node [style=green vertex] (77) at (-1, 0) {};
		\node [style=red vertex] (78) at (-1, 0.75) {};
		\node [style=green vertex] (79) at (0, 0.75) {$\alpha_0$};
		\node [style=green vertex] (81) at (2, 0.75) {$\alpha_2$};
		\node [style=green vertex] (82) at (14.5, 0) {};
		\node [style=red vertex] (83) at (14.5, 0.75) {};
		\node [style=Yexp] (85) at (8.5, 0.75) {};
		\node [style=Zexp] (88) at (6, 0.75) {};
		\node [style=Yexp] (89) at (6, 0) {};
		\node [style=hadamard vertex] (90) at (-1, 1.5) {};
		\node [style=hadamard vertex] (91) at (14.5, 1.5) {};
		\node [style=V] (92) at (-1, 2.25) {};
		\node [style=Vdag] (93) at (14.5, 2.25) {};
		\node [style=green vertex] (94) at (0, 2.25) {};
		\node [style=red vertex] (95) at (0, 1.5) {};
		\node [style=red vertex] (96) at (13.5, 1.5) {};
		\node [style=green vertex] (97) at (13.5, 2.25) {};
		\node [style=green vertex] (100) at (3, 1.5) {};
		\node [style=green vertex] (101) at (12.5, 1.5) {};
		\node [style=red vertex] (102) at (12.5, 0.75) {};
		\node [style=red vertex] (103) at (3, 0.75) {};
		\node [style=V] (104) at (1, 0.75) {};
		\node [style=Vdag] (105) at (13.5, 0.75) {};
		\node [style=none] (106) at (1.5, -0.5) {Quantum Circuit};
		\node [style=none] (107) at (15.25, 3) {};
		\node [style=none] (108) at (15.25, -1) {};
        \node [style=none] (109) at (12.5, -0.5) {Tableau};
	\end{pgfonlayer}
	\begin{pgfonlayer}{edgelayer}
  		\draw (3.center) to (9.center);
        \draw (2.center) to (10.center);
        \draw (1.center) to (11.center);
        \draw (0.center) to (12.center);
		\draw [style=hadamard edge] (49.center) to (50.center);
		\draw [style=hadamard edge] (51.center) to (52.center);
		\draw (63) to (48);
		\draw (83) to (82);
		\draw (78) to (77);
		\draw (85) to (48);
		\draw (88) to (37);
		\draw (89) to (37);
		\draw (94) to (95);
		\draw (97) to (96);
		\draw (100) to (103);
		\draw (101) to (102);
		\draw [style=hadamard edge] (107.center) to (51.center);
		\draw [style=hadamard edge] (107.center) to (108.center);
		\draw [style=hadamard edge] (108.center) to (52.center);
	\end{pgfonlayer}
\end{tikzpicture}

%% file: figures/newexample_f.tikz
\begin{tikzpicture}[baseline={(13)}]
	\begin{pgfonlayer}{nodelayer}
        \node [style=none] (q0) at (-2.2, 0) {$3$};
		\node [style=none] (q1) at (-2.2, 0.75) {$2$};
		\node [style=none] (q2) at (-2.2, 1.5) {$1$};
		\node [style=none] (q3) at (-2.2, 2.25) {$0$};
		\node [style=none] (0) at (-2, 0) {};
		\node [style=none] (1) at (-2, 0.75) {};
		\node [style=none] (2) at (-2, 1.5) {};
		\node [style=none] (3) at (-2, 2.25) {};
		\node [style=none] (9) at (15.5, 2.25) {};
		\node [style=none] (10) at (15.5, 1.5) {};
		\node [style=none] (11) at (15.5, 0.75) {};
		\node [style=none] (12) at (15.5, 0) {};
		\node [style=none] (14) at (-2, 0) {};
		\node [style=none] (15) at (2.5, 0.75) {};
		\node [style=none] (16) at (-2, 1.5) {};
		\node [style=none] (17) at (2.5, 2.25) {};
		\node [style=none] (21) at (5, 2.25) {};
		\node [style=none] (22) at (5, 1.5) {};
		\node [style=none] (23) at (5, 0.75) {};
		\node [style=none] (24) at (5, 0) {};
		\node [style=none] (26) at (5, 0) {};
		\node [style=none] (27) at (5, 0.75) {};
		\node [style=none] (28) at (5, 1.5) {};
		\node [style=none] (29) at (5, 2.25) {};
		\node [style=none] (33) at (7.5, 2.25) {};
		\node [style=none] (34) at (7.5, 1.5) {};
		\node [style=none] (35) at (7.5, 0.75) {};
		\node [style=none] (36) at (7.5, 0) {};
		\node [style=none] (38) at (7.5, 0) {};
		\node [style=none] (39) at (7.5, 0.75) {};
		\node [style=none] (40) at (15.5, 1.5) {};
		\node [style=none] (41) at (7.5, 2.25) {};
		\node [style=none] (44) at (15.5, 2.25) {};
		\node [style=none] (46) at (15.5, 0.75) {};
		\node [style=none] (47) at (15.5, 0) {};
		\node [style=Aexp] (48) at (9.25, 1.125) {$-\alpha_3$};
		\node [style=none] (49) at (7.5, 3) {};
		\node [style=none] (50) at (7.5, -1) {};
		\node [style=none] (51) at (10, 3) {};
		\node [style=none] (52) at (10, -1) {};
		\node [style=green vertex] (77) at (-1, 0) {};
		\node [style=red vertex] (78) at (-1, 0.75) {};
		\node [style=green vertex] (79) at (0, 0.75) {$\alpha_0$};
		\node [style=green vertex] (81) at (2, 0.75) {$\alpha_2$};
		\node [style=green vertex] (82) at (14.5, 0) {};
		\node [style=red vertex] (83) at (14.5, 0.75) {};
		\node [style=hadamard vertex] (90) at (-1, 1.5) {};
		\node [style=hadamard vertex] (91) at (14.5, 1.5) {};
		\node [style=V] (92) at (-1, 2.25) {};
		\node [style=Vdag] (93) at (14.5, 2.25) {};
		\node [style=green vertex] (94) at (0, 2.25) {};
		\node [style=red vertex] (95) at (0, 1.5) {};
		\node [style=red vertex] (96) at (13.5, 1.5) {};
		\node [style=green vertex] (97) at (13.5, 2.25) {};
		\node [style=green vertex] (100) at (3, 1.5) {};
		\node [style=green vertex] (101) at (12.5, 1.5) {};
		\node [style=red vertex] (102) at (12.5, 0.75) {};
		\node [style=red vertex] (103) at (3, 0.75) {};
		\node [style=green vertex] (110) at (4, 0.75) {};
		\node [style=red vertex] (111) at (4, 0) {};
		\node [style=green vertex] (112) at (11.5, 0.75) {};
		\node [style=red vertex] (113) at (11.5, 0) {};
		\node [style=green vertex] (114) at (6, 0) {$\alpha_1$};
		\node [style=Xexp] (115) at (8.5, 0.75) {};
		\node [style=Zexp] (116) at (8.5, 0) {};
		\node [style=V] (117) at (5, 0) {};
		\node [style=Vdag] (118) at (10.5, 0) {};
		\node [style=V] (119) at (1, 0.75) {};
		\node [style=Vdag] (120) at (13.5, 0.75) {};
		\node [style=none] (121) at (2.5, -0.5) {Quantum Circuit};
        \node [style=none] (122) at (12.5, -0.5) {Tableau};
		\node [style=none] (123) at (15.25, 3) {};
		\node [style=none] (124) at (15.25, -1) {};
	\end{pgfonlayer}
	\begin{pgfonlayer}{edgelayer}
   		\draw (3.center) to (9.center);
        \draw (2.center) to (10.center);
        \draw (1.center) to (11.center);
        \draw (0.center) to (12.center);
		\draw [style=hadamard edge] (49.center) to (50.center);
		\draw [style=hadamard edge] (51.center) to (52.center);
		\draw (83) to (82);
		\draw (78) to (77);
		\draw (94) to (95);
		\draw (97) to (96);
		\draw (100) to (103);
		\draw (101) to (102);
		\draw (110) to (111);
		\draw (112) to (113);
		\draw (48) to (115);
		\draw (116) to (48);
		\draw [style=hadamard edge] (123.center) to (51.center);
		\draw [style=hadamard edge] (52.center) to (124.center);
		\draw [style=hadamard edge] (124.center) to (123.center);
	\end{pgfonlayer}
\end{tikzpicture}

%% file: figures/newexample_g.tikz
\begin{tikzpicture}[baseline={(13)}]
	\begin{pgfonlayer}{nodelayer}
        \node [style=none] (q0) at (-2.2, 0) {$3$};
		\node [style=none] (q1) at (-2.2, 0.75) {$2$};
		\node [style=none] (q2) at (-2.2, 1.5) {$1$};
		\node [style=none] (q3) at (-2.2, 2.25) {$0$};
		\node [style=none] (0) at (-2, 0) {};
		\node [style=none] (1) at (-2, 0.75) {};
		\node [style=none] (2) at (-2, 1.5) {};
		\node [style=none] (3) at (-2, 2.25) {};
		\node [style=none] (9) at (15.5, 2.25) {};
		\node [style=none] (10) at (15.5, 1.5) {};
		\node [style=none] (11) at (15.5, 0.75) {};
		\node [style=none] (12) at (15.5, 0) {};
		\node [style=none] (14) at (-2, 0) {};
		\node [style=none] (15) at (2.5, 0.75) {};
		\node [style=none] (16) at (-2, 1.5) {};
		\node [style=none] (17) at (2.5, 2.25) {};
		\node [style=none] (21) at (5, 2.25) {};
		\node [style=none] (22) at (5, 1.5) {};
		\node [style=none] (23) at (5, 0.75) {};
		\node [style=none] (24) at (5, 0) {};
		\node [style=none] (26) at (5, 0) {};
		\node [style=none] (27) at (5, 0.75) {};
		\node [style=none] (28) at (5, 1.5) {};
		\node [style=none] (29) at (5, 2.25) {};
		\node [style=none] (33) at (7.5, 2.25) {};
		\node [style=none] (34) at (7.5, 1.5) {};
		\node [style=none] (35) at (7.5, 0.75) {};
		\node [style=none] (36) at (7.5, 0) {};
		\node [style=none] (38) at (7.5, 0) {};
		\node [style=none] (39) at (7.5, 0.75) {};
		\node [style=none] (40) at (15.5, 1.5) {};
		\node [style=none] (41) at (7.5, 2.25) {};
		\node [style=none] (44) at (15.5, 2.25) {};
		\node [style=none] (46) at (15.5, 0.75) {};
		\node [style=none] (47) at (15.5, 0) {};
		\node [style=none] (51) at (8.75, 3) {};
		\node [style=none] (52) at (8.75, -1) {};
		\node [style=green vertex] (77) at (-1, 0) {};
		\node [style=red vertex] (78) at (-1, 0.75) {};
		\node [style=green vertex] (79) at (0, 0.75) {$\alpha_0$};
		\node [style=green vertex] (81) at (2, 0.75) {$\alpha_2$};
		\node [style=green vertex] (82) at (14.5, 0) {};
		\node [style=red vertex] (83) at (14.5, 0.75) {};
		\node [style=hadamard vertex] (90) at (-1, 1.5) {};
		\node [style=hadamard vertex] (91) at (14.5, 1.5) {};
		\node [style=V] (92) at (-1, 2.25) {};
		\node [style=Vdag] (93) at (14.5, 2.25) {};
		\node [style=green vertex] (94) at (0, 2.25) {};
		\node [style=red vertex] (95) at (0, 1.5) {};
		\node [style=red vertex] (96) at (13.5, 1.5) {};
		\node [style=green vertex] (97) at (13.5, 2.25) {};
		\node [style=green vertex] (100) at (3, 1.5) {};
		\node [style=green vertex] (101) at (12.5, 1.5) {};
		\node [style=red vertex] (102) at (12.5, 0.75) {};
		\node [style=red vertex] (103) at (3, 0.75) {};
		\node [style=green vertex] (110) at (4, 0.75) {};
		\node [style=red vertex] (111) at (4, 0) {};
		\node [style=green vertex] (112) at (11.5, 0.75) {};
		\node [style=red vertex] (113) at (11.5, 0) {};
		\node [style=green vertex] (114) at (6, 0) {$\alpha_1$};
		\node [style=V] (117) at (5, 0) {};
		\node [style=Vdag] (118) at (10.5, 0) {};
		\node [style=hadamard vertex] (119) at (6, 0.75) {};
		\node [style=hadamard vertex] (120) at (10.5, 0.75) {};
		\node [style=green vertex] (121) at (7, 0.75) {};
		\node [style=green vertex] (122) at (9.5, 0.75) {};
		\node [style=red vertex] (123) at (9.5, 0) {};
		\node [style=red vertex] (124) at (7, 0) {};
		\node [style=green vertex] (125) at (8, 0) {$-\alpha_3$};
		\node [style=V] (126) at (1, 0.75) {};
		\node [style=Vdag] (127) at (13.5, 0.75) {};
		\node [style=none] (128) at (15.25, 3) {};
		\node [style=none] (129) at (15.25, -1) {};
		\node [style=none] (130) at (3.75, -0.5) {Quantum Circuit};
		\node [style=none] (131) at (12, -0.5) {Tableau};
	\end{pgfonlayer}
	\begin{pgfonlayer}{edgelayer}
  		\draw (3.center) to (9.center);
        \draw (2.center) to (10.center);
        \draw (1.center) to (11.center);
        \draw (0.center) to (12.center);
		\draw [style=hadamard edge] (51.center) to (52.center);
		\draw (83) to (82);
		\draw (78) to (77);
		\draw (94) to (95);
		\draw (97) to (96);
		\draw (100) to (103);
		\draw (101) to (102);
		\draw (110) to (111);
		\draw (112) to (113);
		\draw (121) to (124);
		\draw (122) to (123);
		\draw [style=hadamard edge] (51.center) to (128.center);
		\draw [style=hadamard edge] (128.center) to (129.center);
		\draw [style=hadamard edge] (129.center) to (52.center);
	\end{pgfonlayer}
\end{tikzpicture}

%% file: sections/experiments.tex
We investigated the performance of the proposed methods on both architectures with all-to-all connectivity (refereed to as complete architectures) and restricted architectures, namely those available in the IBM Software stack\footnote{Our implementation can be found on Github: \url{https://github.com/daehiff/pauliopt/tree/new_methods}. 
}. In our experiments, we targeted the following IBM backends: \textit{quito} (5 qubits),  \textit{nairobi} (7 qubits), \textit{guadalupe} (16 qubits), \textit{mumbai} (27 qubits), and \textit{ithaca} (65 qubits). 

All experiments were run using a process pool where each combination of circuit and synthesis strategy was distributed as tasks in a multi-threaded manner. Each synthesis strategy ran completely serially with no other innate parallelism.

For the randomized experiments, we synthesized $20$ random Pauli polynomials with increasing gadget count for each architecture. The leg count was uniformly sampled from $\{1..q\}$, where $q$ is the size of the given architecture. The leg positions were randomly selected and its Pauli letter was uniformly sampled from $\{X,Y,Z\}$. The gadget angle was uniformly sampled from $\{\pi, \frac{\pi}{2}, \frac{\pi}{4}, \frac{\pi}{8}, \frac{\pi}{16}\}$. 

Additionally, we synthesized the molecules present in pytket's UCCSD test set. These correspond to Hamiltonians used to estimate the ground state energy of small molecules and were used as benchmarks in \cite{Sivarajah2020, cowtan2020generic}.

We compare the proposed methods against Paulihedral \cite{li2022paulihedral}  and the UCCSD pair-wise and set-based synthesis methods in~\cite{cowtan2019phase}. As a baseline, we included a naive Steiner tree-based decomposition which did not use any reordering. An additional routing step utilizing pytket~\cite{Sivarajah2020} is used for methods that are not architecture-aware. To compare apples with apples, all circuits are synthesized to the gateset $\{\text{CNOT}, U3\}$ and evaluated on CNOT count, circuit depth, and CNOT depth. 

Since the gadget ordering found by our algorithm affects the Trotter error, we also calculated the Trotter error of all methods for random circuits and two molecules: H2 (in 631g with parity encoding) and H4 (in sto3g with parity encoding).

\subsection{Random Pauli polynomials}
\begin{figure*}[h]
    \begin{subfigure}{\textwidth} 
    \centering\includegraphics[width=\textwidth]{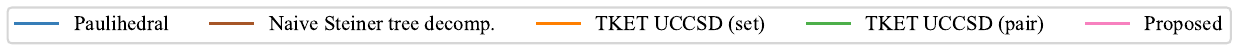}
    \end{subfigure}
    \hfill
        \begin{subfigure}{\textwidth}
    	    \centering
    		\includegraphics[width=\textwidth,trim={8cm 0cm 7cm 1cm},clip]{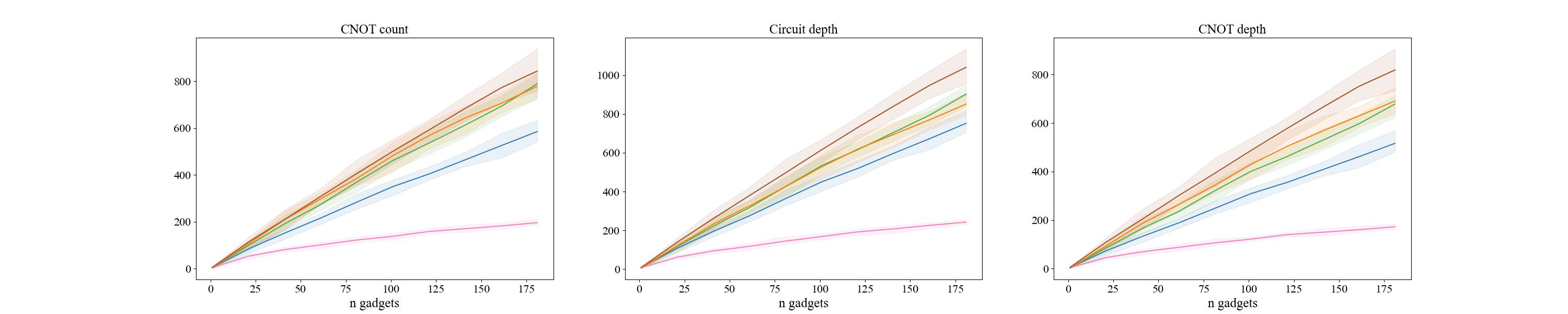}%
    		\caption{Quito (5 Qubits)}
    	\end{subfigure}
    \hfill
     	\begin{subfigure}{\textwidth}
    	    \centering
    		\includegraphics[width=\textwidth,trim={8cm 0cm 7cm 1cm},clip]{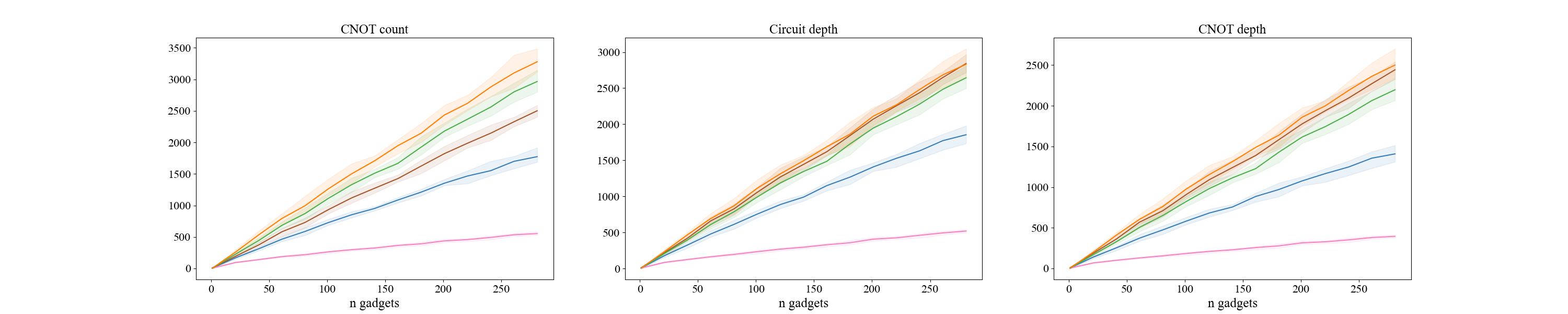}
    		\caption{Nairobi (7 Qubits)}
    	\end{subfigure}
     \hfill
    	\begin{subfigure}{\textwidth}
    	    \centering
    		\includegraphics[width=\textwidth,trim={8cm 0cm 7cm 1cm},clip]{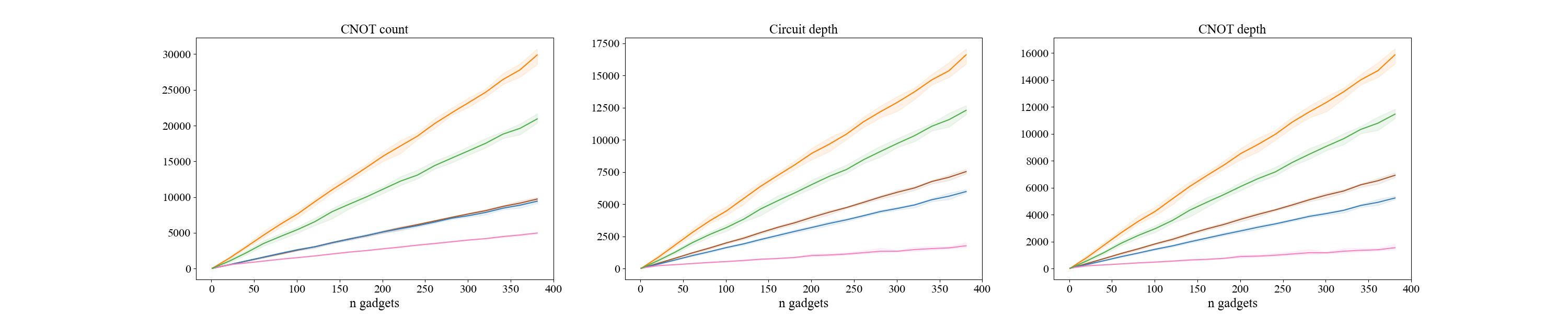}
    		\caption{Guadalupe (16 Qubits)}
    	\end{subfigure}
     \hfill
    	\begin{subfigure}{\textwidth}
    	    \centering
    		\includegraphics[width=\textwidth,trim={8cm 0cm 7cm 1cm},clip]{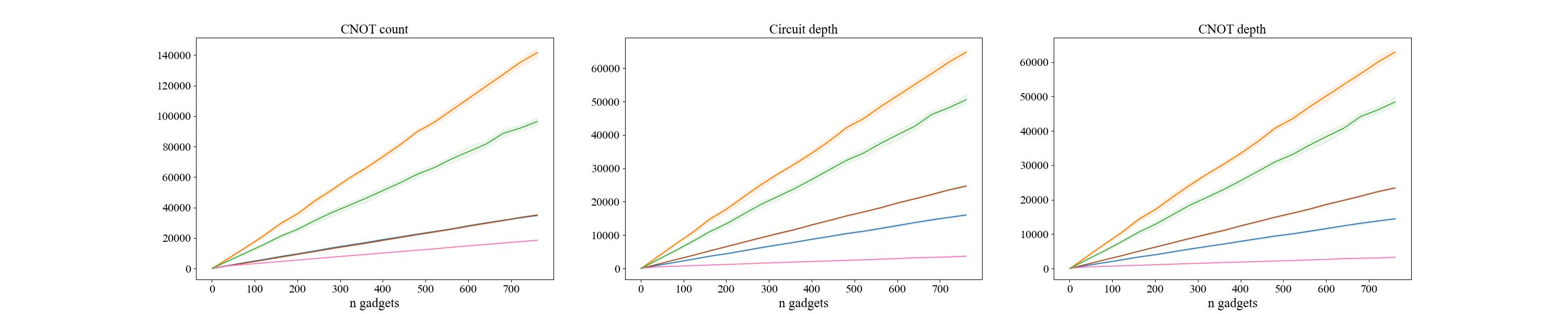}
    		\caption{Mumbai (27 Qubits)}
    	\end{subfigure}
     \hfill
    	\begin{subfigure}{\textwidth}
    	    \centering
    		\includegraphics[width=\textwidth,trim={8cm 0cm 7cm 1cm},clip]{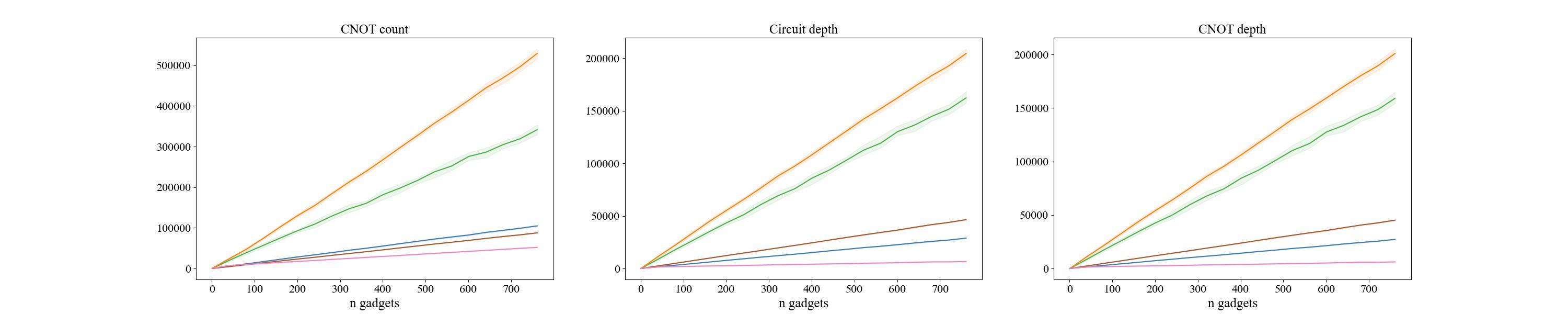}
    		\caption{Ithaca (65 Qubits)}
    	\end{subfigure}
    \caption{Comparison of examined synthesis methods on CNOT count, circuit depth and 2-qubit gate depth for different IBM quantum computers.}\label{fig:comparison_restricted}
\end{figure*}

\begin{figure*}[bt]
    \begin{subfigure}{\textwidth} 
    \centering
    \includegraphics[width=\textwidth]{figures/legend.pdf}
    \end{subfigure}
    \hfill
        \begin{subfigure}{\textwidth}
    	    \centering
            \includegraphics[width=\textwidth,trim={8cm 0cm 7cm 1cm},clip]{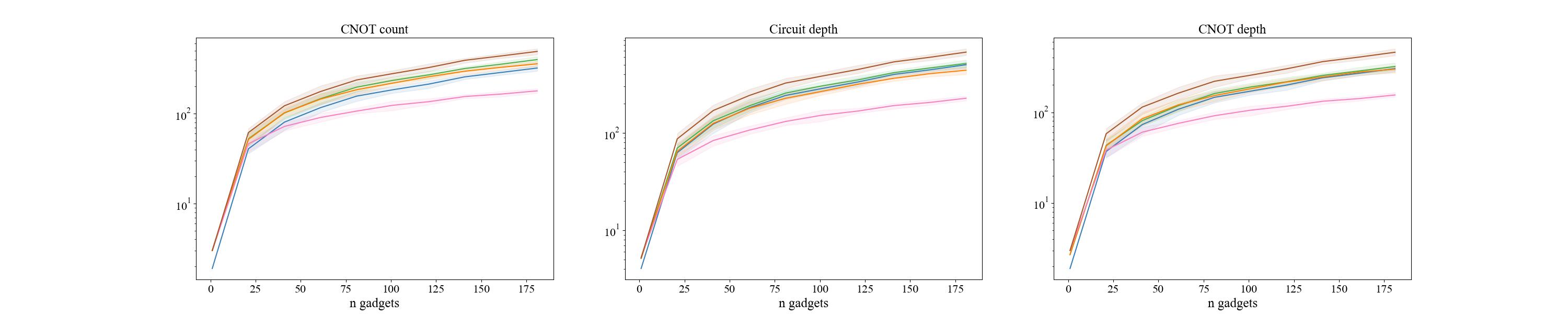}%
    		\caption{Complete (5 Qubits)}
    	\end{subfigure}
     \hfill
     	\begin{subfigure}{\textwidth}
    	    \centering
    		\includegraphics[width=\textwidth,trim={8cm 0cm 7cm 1cm},clip]{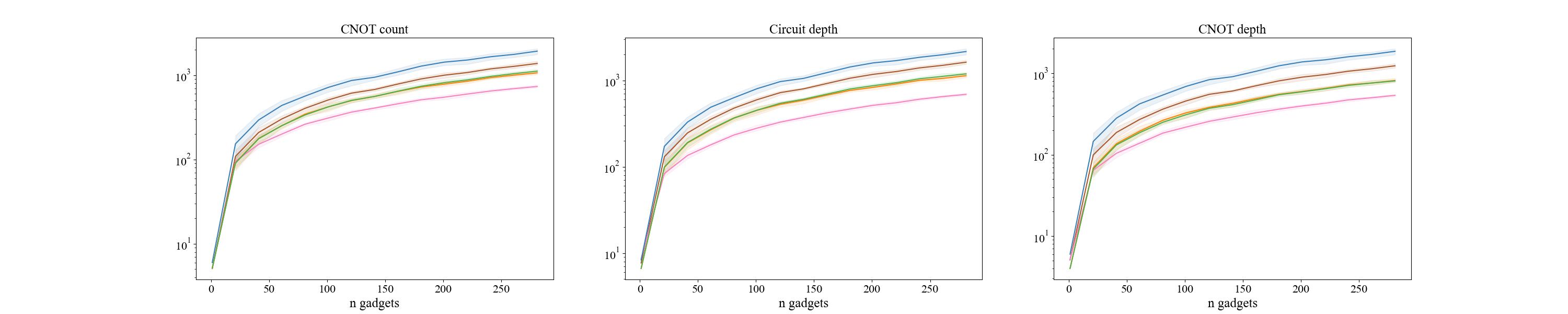}
    		\caption{Complete (7 Qubits)}
    	\end{subfigure}
     \hfill
        \begin{subfigure}{\textwidth}
    		\includegraphics[width=\textwidth,trim={8cm 0cm 7cm 1cm},clip]{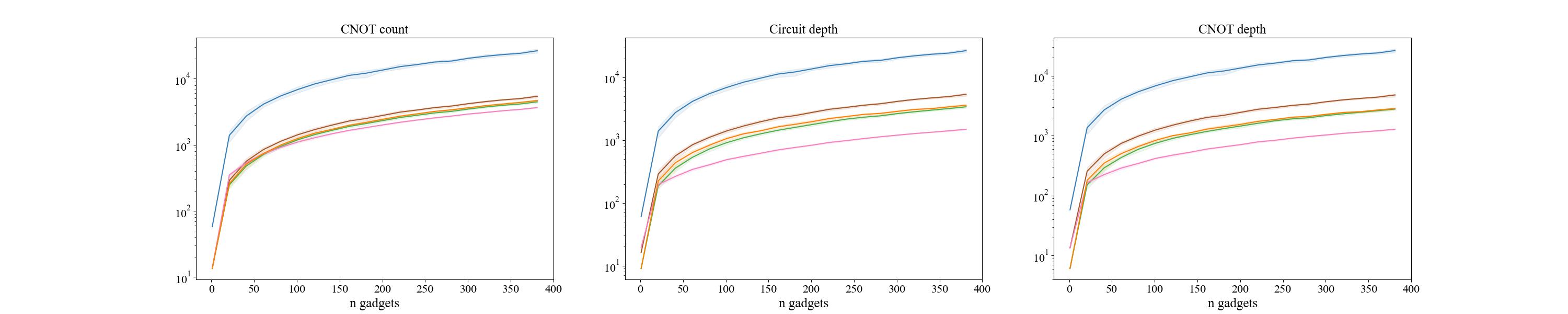}
    		\caption{Complete (16 Qubits)}
    	\end{subfigure}
     \hfill
        \begin{subfigure}{\textwidth}
    		\includegraphics[width=\textwidth,trim={8cm 0cm 7cm 1cm},clip]{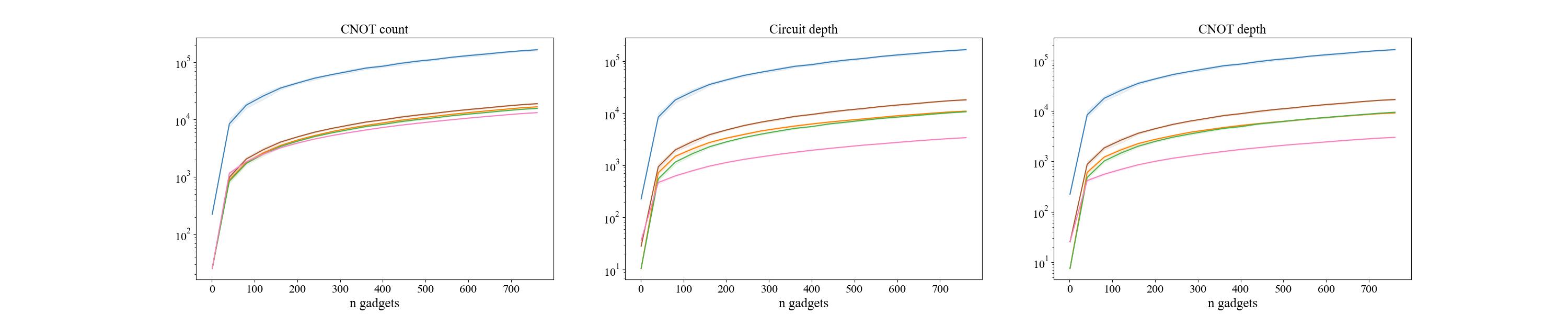}
    		\caption{Complete (27 Qubits)}
    	\end{subfigure}
     \hfill
        \begin{subfigure}{\textwidth}
    		\includegraphics[width=\textwidth,trim={8cm 0cm 7cm 1cm},clip]{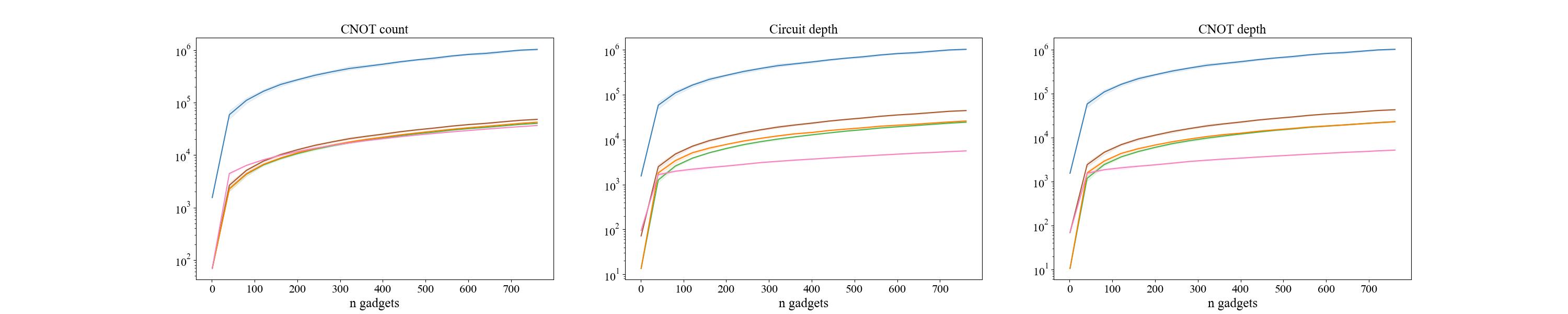}
    		\caption{Complete (65 Qubits)}
    	\end{subfigure}
	\caption{The performance of the different synthesis methods on different metric for different size architectures with all-to-all connectivity (complete).}\label{fig:comparison_complete}
\end{figure*}

We present the average CNOT count, circuit depth, and two-qubit gate depth of the synthesized circuit in \autoref{fig:comparison_restricted} and \autoref{fig:comparison_complete}, for restricted and complete architectures respectively.
For both cases, the proposed algorithm improves across the board for deep circuits (>50 gadgets). We hypothesize that this is a consequence of collecting operations in a Clifford tableau, which avoids the construction of the full CNOT ladder (i.e. sequence of CNOTs) in gadget synthesis as we propagate the second half of the ladder through the remainder of the polynomial. Thus, subsequent gadgets can now produce transverse two-qubit gates. 
Interestingly, we see that for shallow circuits (<50 gadgets), the proposed method performs worse on CNOT count, we attribute this to the greedy choosing of rows with large numbers of $I$ legs, effectively routing rotations through long chains of $I$. Additionally, the proposed algorithm uses a naive initial placement of the qubits while both Paulihedral and TKET are able to find a better qubit mapping as a starting point for synthesis.

\subsection{UCCSD molecular ansatzes}\label{subsec:results-uccsd}
Comparison of the three metrics for the various examined algorithms can be found in \autoref{table:uccs-cnot}, \autoref{table:uccs-depth}, and \autoref{table:uccs-2qdepth}. In this experiment, we synthesized each molecule for the smallest IBM device that would fit the circuit. We applied a time-out of 60 minutes but we found that in some cases the threads had been suspended resulting in incorrect runtimes. For a fair comparison, we decided to only present the molecules for which all methods finished and to omit runtime results.

From the tables, we can see that the proposed algorithm shows a clear improvement over the other examined methods; the only exception is the three smallest molecules. We hypothesize that this is, once again, due to the lack of qubit mapping and the greedy heuristic .

\subsection{Evaluation of the Trotter error}
As the chosen ordering of Pauli gadgets is determined heuristically, the new order may increase the resulting Trotter error. Hence, we investigated the unitary overlap of the exact time evolution operator $U_{\text{exact}} = \exp \left( -i \frac{\theta}{2} H\right)$  and the synthesized circuit $U_{\text{circuit}}$, as in  \autoref{eq:unitary_overlap}:
\begin{equation}\label{eq:unitary_overlap}
    \text{overlap} = \bra{0} U_{\text{exact}} U_{\text{circuit}}^\dagger \ket{0}
\end{equation}
We used $20$ randomly generated Pauli polynomials with six qubits and 160 gadgets. Each gadget had uniformly distributed legs across the six qubits with a type in $\left[X, Y, Z\right]$. The rotation angle was uniformly sampled from: $\left[ \nicefrac{\pi}{32}, \nicefrac{\pi}{64}, \nicefrac{\pi}{128}\right]$. Additionally, we performed this investigation on molecules \texttt{H2\_P\_631g} and \texttt{H4\_P\_sto3g} from pytket's UCCSD test set~\cite{cowtan2020generic}. For those molecules, we again picked the angles of the individual gadgets at random from the set $\left[ \nicefrac{\pi}{32}, \nicefrac{\pi}{64}, \nicefrac{\pi}{128}\right]$ to facilitate a fair comparison among the experiments.

\begin{figure}
    \centering\includegraphics[width=\linewidth]{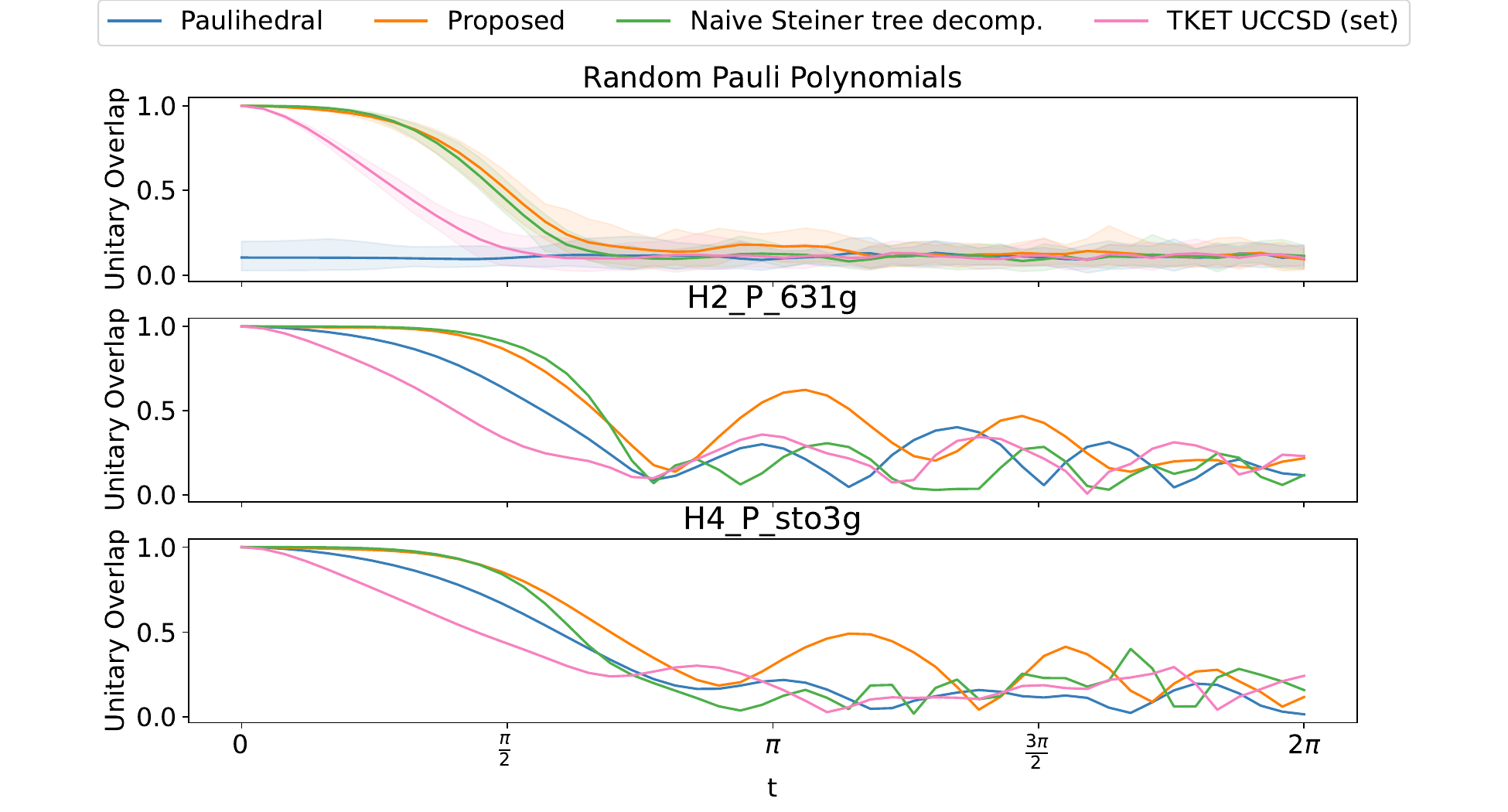}
    \caption{Unitary overlap between the exact time evolution operator and the synthesized circuit for 20 random Pauli polynomials (aggregated) and the molecules \texttt{H2\_P\_631g} and \texttt{H4\_P\_sto3g}.}
    \label{fig:plots_fid}
\end{figure}

We examined the overlap for all cases over multiple timesteps in $\left[0:2\pi\right]$, and the results are plotted in \autoref{fig:plots_fid}. We omit TKET UCCSD (pair) from the figure as it does not change the ordering of the Pauli gadgets and the Trotter error would be equal to that of naive synthesis.
Across the various cases, we see that the proposed method is comparable to default Trotterization. Still, the best possible performance of each algorithm depends on the molecule's structure.
\csvstyle{myTableStyle}{tabular=|l|l|c|c|r|r|r|r|r|,
        table head=\hline \bfseries Molecule. & \bfseries Backend. & \bfseries \# Qubits. & \bfseries Naive & \bfseries Paulihedral & \bfseries TKET (set) & \bfseries TKET (pair) & \bfseries \bfseries Proposed \\\hline\hline,
        late after line=\\\hline,
        head to column names,
        respect underscore=true
        }
\begin{table*}[t]
\centering
\begin{filecontents*}{figures/molecules_new_renamed.txt}
,name,backend,numqubits,ngadgets,naivecx,naiveu3,naivedepth,naiveqqdepth,paulihedralcx,paulihedralu3,paulihedraldepth,paulihedralqqdepth,tketuccspaircx,tketuccspairu3,tketuccspairdepth,tketuccspairqqdepth,tketuccssetcx,tketuccssetu3,tketuccssetdepth,tketuccssetqqdepth,paulioptsteinercliffordcx,paulioptsteinercliffordu3,paulioptsteinerclifforddepth,paulioptsteinercliffordqqdepth
263,H2_P_sto3g,quito,2,4,\textbf{4},8,9,4,\textbf{4},8,9,4,\textbf{4},9,8,4,\textbf{4},11,9,4,5,\textbf{4},\textbf{4},\textbf{3}
19,H2_BK_sto3g,quito,4,12,64,37,85,64,52,37,76,52,38,45,68,38,\textbf{24},26,\textbf{42},\textbf{24},35,37,47,35
237,H2_JW_sto3g,quito,4,12,62,51,87,62,64,49,87,64,38,57,75,38,\textbf{36},27,55,\textbf{36},38,39,\textbf{53},38
218,H2_P_631g,nairobi,6,158,1 190,557,1 461,1 172,1 056,608,1 150,872,1 086,471,1 143,893,671,246,646,517,\textbf{433},\textbf{366},\textbf{444},\textbf{332}
181,H4_P_sto3g,nairobi,6,164,1 214,578,1 497,1 196,1 072,633,1 170,884,1 125,490,1 181,923,709,252,680,549,\textbf{471},\textbf{392},\textbf{466},\textbf{344}
275,H2_JW_631g,guadalupe,8,84,894,361,908,739,783,368,783,619,691,392,815,560,458,190,496,381,\textbf{414},\textbf{280},\textbf{339},\textbf{272}
95,H2_BK_631g,guadalupe,8,84,1 042,292,979,822,862,345,796,607,702,352,786,572,435,190,423,330,\textbf{443},\textbf{293},\textbf{369},\textbf{301}
72,H4_BK_sto3g,guadalupe,8,160,2 042,584,1 814,1 518,1 697,679,1 667,1 316,1 613,735,1 669,1 231,1 532,370,1 316,1 113,\textbf{878},\textbf{549},\textbf{656},\textbf{523}
223,H4_JW_sto3g,guadalupe,8,160,1 896,698,1 851,1 532,1 505,737,1 449,1 112,1 686,727,1 590,1 159,1 175,310,950,815,\textbf{883},\textbf{567},\textbf{677},\textbf{552}
162,NH_BK_sto3g,guadalupe,12,640,12 660,2 635,10 742,9 552,7 231,2 778,6 494,5 250,11 164,3 119,9 463,7 791,9 299,1 496,6 685,5 937,\textbf{4 788},\textbf{3 176},\textbf{2 986},\textbf{2 433}
169,LiH_BK_sto3g,guadalupe,12,640,11 210,2 565,9 659,8 476,8 681,3 017,6 812,5 488,10 980,2 908,9 690,8 010,8 682,1 411,6 213,5 516,\textbf{4 250},\textbf{2 835},\textbf{2 760},\textbf{2 261}
42,LiH_JW_sto3g,guadalupe,12,640,10 108,2 795,9 219,7 959,7 858,2 969,6 926,5 672,8 149,2 923,7 813,6 005,5 793,1 220,4 278,3 665,\textbf{4 331},\textbf{2 911},\textbf{2 944},\textbf{2 385}
176,NH_JW_sto3g,guadalupe,12,640,12 880,2 805,11 390,10 130,7 645,2 937,6 480,5 195,9 089,2 929,8 612,6 830,5 779,1 216,4 338,3 712,\textbf{4 884},\textbf{3 237},\textbf{3 145},\textbf{2 608}
212,H2O_P_sto3g,guadalupe,12,1 085,19 218,5 606,16 365,14 429,12 381,5 745,9 627,7 893,20 089,4 216,15 340,13 450,17 527,1 937,11 470,10 570,\textbf{8 077},\textbf{4 492},\textbf{4 655},\textbf{4 015}
282,BeH2_P_sto3g,guadalupe,12,1 085,19 950,5 735,16 977,15 012,12 405,5 913,10 179,8 241,19 761,4 482,15 450,13 435,18 717,1 956,11 822,10 994,\textbf{8 482},\textbf{4 641},\textbf{5 053},\textbf{4 372}
25,CH2_P_sto3g,guadalupe,12,2 109,38 344,10 887,32 899,29 062,22 345,11 069,18 201,14 602,40 180,8 609,31 610,27 773,32 667,3 548,21 278,19 596,\textbf{14 909},\textbf{8 559},\textbf{9 049},\textbf{7 740}
38,H2O_JW_sto3g,guadalupe,14,1 000,19 668,4 387,18 907,16 941,12 816,4 558,11 074,9 112,17 362,4 643,15 770,12 912,11 193,2 039,8 295,7 205,\textbf{7 967},\textbf{5 034},\textbf{4 420},\textbf{3 683}
193,H2O_BK_sto3g,guadalupe,14,1 000,23 986,4 488,20 285,18 572,14 417,5 117,11 989,10 000,23 090,5 091,19 074,16 404,19 585,2 572,13 347,12 128,\textbf{8 675},\textbf{5 281},\textbf{4 616},\textbf{3 877}
135,CH2_BK_sto3g,guadalupe,14,1 488,34 352,6 549,29 604,26 998,20 890,7 556,17 397,14 388,35 106,7 700,27 314,23 437,35 847,3 845,22 198,20 581,\textbf{12 919},\textbf{8 278},\textbf{7 361},\textbf{6 182}
109,BeH2_JW_sto3g,guadalupe,14,1 488,27 602,6 525,25 671,22 751,20 315,7 072,16 772,13 953,25 952,6 879,22 939,18 938,19 296,2 818,12 702,11 345,\textbf{12 181},\textbf{7 821},\textbf{7 061},\textbf{5 908}
283,BeH2_BK_sto3g,guadalupe,14,1 488,30 786,6 777,26 175,23 501,22 887,7 618,17 404,14 489,35 030,7 632,27 319,23 515,34 138,3 738,21 184,19 583,\textbf{12 419},\textbf{7 910},\textbf{6 749},\textbf{5 628}
146,CH2_JW_sto3g,guadalupe,14,1 488,31 244,6 537,28 816,25 891,19 913,7 001,16 954,14 032,25 863,6 877,22 688,18 618,17 744,2 783,11 872,10 520,\textbf{13 433},\textbf{8 232},\textbf{7 525},\textbf{6 398}
244,H4_P_631g,guadalupe,14,2 912,58 612,16 670,49 277,44 163,36 203,17 899,28 669,23 662,61 530,12 968,47 166,41 739,38 959,4 527,25 782,23 556,\textbf{24 224},\textbf{13 355},\textbf{12 722},\textbf{10 984}
239,H4_JW_631g,mumbai,16,1 440,26 382,6 286,25 820,23 012,21 083,6 703,17 934,15 236,25 416,6 694,23 024,18 842,15 012,2 829,10 752,9 246,\textbf{11 727},\textbf{7 329},\textbf{6 390},\textbf{5 267}
54,H4_BK_631g,mumbai,16,1 440,29 146,5 863,26 139,23 576,23 136,66 74,16 739,14 038,26 366,6 644,22 429,18 558,22 365,3 119,15 712,14 132,\textbf{13 852},\textbf{8 188},\textbf{5 707},\textbf{4 585}
184,NH3_JW_sto3g,mumbai,16,2 340,46 902,10 240,45 910,41 306,33 483,10 915,28 846,24 436,49 657,10 986,40 676,34 266,35 769,4 720,23 085,20 783,\textbf{23 182},\textbf{13 827},\textbf{9 734},\textbf{7 937}
209,NH3_BK_sto3g,mumbai,16,2 340,57 254,10 763,49 835,45 843,35 990,12 149,28 640,23 905,63 328,12 447,46 464,40 274,56 862,5 921,34 055,31 523,\textbf{25 280},\textbf{14 920},\textbf{10 121},\textbf{8 400}
154,HCl_JW_sto3g,mumbai,20,684,19 854,2 975,16 055,14 686,9 430,2 889,8 452,7 195,13 108,3 423,11 069,9 104,9 780,1 729,7 829,6 785,\textbf{5 807},\textbf{2 956},\textbf{3 029},\textbf{2 677}
152,HCl_BK_sto3g,mumbai,20,684,21 746,3 164,18 055,16 773,9 867,3 258,7 901,6 623,13 499,3 559,11 116,9 165,10 571,1 786,8 449,7 475,\textbf{6 306},\textbf{3 223},\textbf{3 113},\textbf{2 720}
\end{filecontents*}
\csvreader[myTableStyle]{figures/molecules_new_renamed.txt}{}%
{\name & \backend & \numqubits & \naivecx & \paulihedralcx  & \tketuccssetcx & \tketuccspaircx & \paulioptsteinercliffordcx }
\caption{CNOT Count for different synthesis methods UCCSD molecule benchmark set on different IBM architectures. The smallest value in the row is in bold font.}
\label{table:uccs-cnot}
\end{table*}

\begin{table*}[b]
\centering
\csvreader[myTableStyle]{figures/molecules_new_renamed.txt}{}%
{\name & \backend & \numqubits & \naivedepth & \paulihedraldepth  & \tketuccssetdepth & \tketuccspairdepth & \paulioptsteinerclifforddepth }
\caption{Circuit depth for different synthesis methods UCCSD molecule benchmark set on different IBM architectures. The smallest value in the row is in bold font.}
\label{table:uccs-depth}
\end{table*}

\begin{table*}[b]
\centering
\csvreader[myTableStyle]{figures/molecules_new_renamed.txt}{}%
{\name & \backend & \numqubits &\naiveqqdepth & \paulihedralqqdepth  & \tketuccssetqqdepth & \tketuccspairqqdepth & \paulioptsteinercliffordqqdepth }
\caption{2 qubit gate depth for different synthesis methods UCCSD molecule benchmark set on different IBM architectures. The smallest value in the row is in bold font.}
\label{table:uccs-2qdepth}
\end{table*}

%% file: sections/conclusion.tex
In this work, we introduce a novel method for architecture-aware Pauli polynomial synthesis, which propagates synthesized gates through the remainder of the polynomial. As a consequence, we introduce an alternative lexicographical grouping methodology that emerges from the structure of the proposed algorithm. We additionally describe an extension to the architecture-aware synthesis introduced in \cite{winderl2023architectureaware} that allows Clifford tableau synthesis up-to-permutation. We observe that the introduced method outperforms other examined methods on both restricted and complete architectures and, for the small set of test cases observed, does not seem to introduce significant Trotter error when compared with similar algorithms. 

Although investigation of the Trotter error associated with the proposed method is preliminary, we note that the error introduced by the noisy gates in the resulting circuits of alternative methods vastly outweighs the potential difference in Trotter error in cases where the proposed method worsens Trotter error. Considering the scale of improvement in gate counts and depth as observed with our method, reducing Trotter error will become a factor in fault tolerant devices.

Nevertheless, there are several areas for future improvement. One key weakness of our introduced algorithm is the cost function for pivot selection; for shallow regions, it removes interactions inefficiently. This means that the final recursions in the algorithm likely introduce unnecessary two-qubit operations. This can be addressed by improving or tuning the cost function or by introducing methods such as reverse traversal~\cite{Li2019} to improve the initial qubit mapping. The current algorithm also does not find a good qubit placement during synthesis, incorporating a scheme similar to PermRowCol may be further improve its overall performance.

We additionally note that the time to solution for the proposed method is not ideal, taking up to 30 minutes on a local workstation for the largest, deepest circuits in our experiments. Although this can be partially attributed to unoptimized python code, algorithmic improvements may be necessary for future larger devices. Methodologies such as `parallel-in-time' may be interesting to investigate if the cost of propagated Clifford tableaus through synthesized regions is low and does not impact performance significantly. The current algorithm can also beextended to support the synthesis of higher-order symmetric product formulas mentioned in \cite{gui2021term} or other higher-order decompositions 
\cite{Childs2021, Ostmeyer_2023, Low2023}. Additionally, the algorithm can be represented as row additions in a $2q\times g$ matrix, where $q$ is the number of qubits and $g$ the number of Pauli gadgets, which suggests that an efficient GPU implementation is possible.

In summary, the architecture-aware synthesis of Pauli gadgets offers a powerful alternative to existing methods to synthesize Pauli-based product formulas on restricted architectures; architecture-conforming algorithms reduce routing overhead tremendously, a key step to enable effective quantum hardware usage in the near term.